\documentclass{aa}

\usepackage{graphicx}
\usepackage[normalem]{ulem} 
\usepackage{txfonts}
\usepackage{amssymb}
\usepackage{enumerate}
\usepackage{subcaption}

\usepackage[colorlinks=true,allcolors=blue]{hyperref}%
\renewcommand{\vec}[1]{{\mathbf #1}}

\newcommand{\llnr}[1]{{\bf \color{magenta}{[]}} \color{black}} 

\usepackage{xcolor}
\usepackage{soul}
\definecolor{mydarkgreen}{RGB}{0,100,0}

\usepackage{academicons}
\definecolor{orcidlogocol}{HTML}{A6CE39}

\newcommand{\orcid}[1]{\href{https://orcid.org/#1}{\textcolor[HTML]{A6CE39}{\aiOrcid}}}

\begin{document}

 \title{CME propagation in the dynamically coupled space weather tool: COCONUT + EUHFORIA }

 \author{L. Linan\href{https://orcid.org/0000-0002-4014-1815}{\includegraphics[scale=0.05]{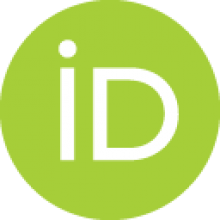}}\inst{1} \and T. Baratashvili\inst{1}  \and A. Lani\inst{1}  \and B. Schmieder 
 \href{https://orcid.org/0000-0003-3364-9183}{\includegraphics[scale=0.05]{orcid-ID.png}}\inst{1,4,5} \and M. Brchnelova\inst{1}  \and J. H. Guo\href{https://orcid.org/0000-0002-4205-5566}{\includegraphics[scale=0.05]{orcid-ID.png}}\inst{1,5} \and S. Poedts\href{https://orcid.org/0000-0002-1743-0651}{\includegraphics[scale=0.05]{orcid-ID.png}}\inst{1,2}}
 
 \institute{Centre for mathematical Plasma-Astrophysics, Department of Mathematics, KU Leuven, Celestijnenlaan 200B, 3001 Leuven, Belgium \\
  \and Institute of Physics, University of Maria Curie-Sk{\l}odowska, ul.\ Radziszewskiego 10, 20-031 Lublin, Poland \\
    \and LIRA, Observatoire de Paris, Université PSL, CNRS, Sorbonne Université, Université de Paris, 5 place Jules Janssen, 92190 Meudon, France\\
  \and University of Glasgow, School of Physics and Astronomy, Glasgow, G128QQ, Scotland \\
  \and School of Astronomy and Space Science and Key Laboratory of Modern Astronomy and Astrophysics, Nanjing University, Nanjing 210023, China }

 \date{Submitted}

 
 \abstract
 {Numerical MHD models like the European Heliospheric Forecasting Information Asset (EUHFORIA) have been developed to predict the arrival time of Coronal Mass Ejections (CMEs) and accelerated high-energy particles. However, in EUHFORIA, transient magnetic structures are injected at 0.1 AU into a background solar wind created from a static solar wind model. This means the inserted CME model is completely independent of the coronal magnetic field, missing all potential interactions between the CME and the solar wind in the corona.}
 {This paper aims to present the time-dependent coupling between the coronal model COolfluid COroNal UnsTructured (COCONUT) and the heliospheric forecasting tool EUHFORIA.{\bf The first attempt of coupling these two simulations should allow us to follow directly the propagation of flux rope from the Sun to Earth.}}
 {We perform six COCONUT simulations where a flux rope is implemented at the solar surface using either the Titov-Démoulin CME model or the Regularized Biot-Savart Laws (RBSL) CME model. At regular intervals, the magnetic field, velocity, temperature, and density of the 2D surface $R_{b}=21.5~\;R_{\odot}$ are saved in boundary files. This series of coupling files is read in a modified version of EUHFORIA to update progressively its inner boundary. After presenting the early stage of the propagation in COCONUT, we examine how the disturbance of the solar corona created by the propagation of flux ropes is transmitted into EUHFORIA. In particular, we consider the thermodynamic and magnetic profiles at L1 and compare them with those obtained at the interface between the two models.}
 {We demonstrate that the properties of the heliospheric solar wind in EUHFORIA are consistent with those in COCONUT, acting as a direct extension of the coronal domain. Moreover, the disturbances initially created from the propagation of flux ropes in COCONUT continue evolving from the corona in the heliosphere to Earth with a smooth transition at the interface between the two simulations. Looking at the profile of magnetic field components at Earth and different distances from the Sun, we also find that the transient magnetic structures have a self-similar expansion in COCONUT and EUHFORIA. However, the amplitude of the profiles depends on the flux rope model used and its properties, thus emphasizing the important role of the initial properties in solar source regions for accurately predicting the impact of CMEs.}
 {The dynamically coupled COCONUT + EUHFORIA model chain constitutes a new space weather forecasting tool that can predict flux-rope CMEs characteristics upon their arrival at L1.} 



 \maketitle
%
\section{Introduction} \label{sec:Introduction}
 
Extreme solar events \citep{Gosling93,Schrijver15} have been listed as one of the most significant risks coming from space, potentially leading to catastrophic events that forecasting assets must be able to predict as accurately as possible to protect humanity. Particles with high energy, called Solar Energetic Particle (SEP), can be accelerated during their journey by the solar wind or by magnetic structures in the interplanetary medium \citep{Reames13}. Flare- and shock-related processes are two of the leading candidates for the efficient acceleration of SEPs observed in situ \citep[see, e.g., reviews by][]{Desai16,Klein17,Vlahos19}.

Magnetic storms are another type of event caused by Coronal Mass Ejections (CMEs) that are expelled from the Sun with speed reaching 1000 to 2000 km/s \citep{Schmieder15}. Fast and wide CMEs are powerful drivers of shockwaves in the corona and interplanetary space that, in turn, can efficiently accelerate particles to high energies and release them in very distant locations from the flare or eruption site \citep[e.g.,][]{Rouillard16,Afanasiev18,Kouloumvakos19,Kouloumvakos22}. Magnetic disturbances on Earth can also be initiated by corotating interaction regions (CIRs), which form when high-speed streams (HSS) emanating from a coronal hole interact with the ambient slow solar wind \citep{Hajra22}. However, the most severe geomagnetic storms arise when the $B_z$ component of the CME’s magnetic field — aligns in the opposite direction to the Earth's magnetopause \citep{Lugaz16}. This opposite alignment facilitates magnetic reconnection between the magnetic fields of these two structures, thereby channelling energy into the Earth's inner magnetosphere \citep{Akasofu81,Dungey61}.

These events can cause catastrophic effects on human infrastructure. Geomagnetically induced currents, created due to space weather events, can enter the power infrastructure and saturate or damage the equipment (e.g. transformers), causing power blackouts over large areas in case of severe storms \citep{Eastwood17}. Additionally, high-energy particles might compromise spacecraft operations, posing a risk to equipment and astronauts' safety \citep{Bothmer07,Baker04}. Understanding the physics of Sun-Earth transients and the mechanisms driving space weather events is crucial for reducing potential risks and associated costs.

In the heliosphere, the properties of Interplanetary Coronal Mass Ejections (ICMEs) are measured at various distances from Earth using a suite of spacecraft such as ACE \citep{Chiu98}, the Parker Solar Probe \citep{Fox16}, the Solar Orbiter \citep{Muller20}, WIND \citep{Harten95}, STEREO A and B \citep{Kaiser07}. While methods exist to reconstruct the 3D structure of ICMEs from multiple viewpoints \citep{Rodari18}, these methods are complex due to the limited number of available observations \citep{Demoulin10}. Consequently, these observations are supplemented by numerical simulations designed to model the properties of the solar wind and/or predict the evolution of the Sun's magnetic structures as they travel to Earth.

Even though some simulation tools \citep[e.g.][]{Manchester04,Mignone07,Lugaz05,Regnault23,Shen07,Shen13}, can be used to track a CME from the Sun to Earth, it is generally preferable to separate the study into different modelling tools due to the varying physical conditions from the solar corona to the heliosphere \citep{Forbes06}. Additionally, having one single large domain requires numerous grid points, leading to a notable slowdown in the wall-clock times of the simulations.

Therefore, among the various tools available, we find a set of numerical simulations specifically designed to reproduce the solar corona's complexities accurately. For instance, the Wang-Sheeley-Arge model \citep[WSA;]{Wang90,McGregor11,van10} is commonly used to create a solar wind from the observations of photospheric magnetograms \citep[e.g.,][]{Verbeke19,Scolini19,Prete23}. Other notable examples include the steady global corona model with turbulence transport and heating by \citet{Usmanov96, Usmanov14, Usmanov18, Chhiber21}, the Wind-Predict-AW code \citep{Reville2020, Reville2021, Parenti2022}, and the Alfvén Wave Solar atmosphere Model \citep[AWSoM;][]{van10,Van15,Sachdeva19} already integrated within the Space Weather Modeling Framework \citep[SWMF;][]{toth12}. A more advanced model describing the ambient plasma density, temperature, and magnetic field of the background corona is provided by the Magneto-hydrodynamics Around a Sphere (MAS) model from the solar surface to $30~\;R_{\odot}$. This model was developed by Predictive Science Inc \citep{Riley11,Lionello08}.

In this work, we will focus on the global 3D MHD coronal plasma solver known as COolfluid COroNal UnsTructured \citep[COCONUT,][]{Perri22,Perri23}. COCONUT's implicit scheme, coupled with its highly scalable parallel architecture, sets it apart from other solvers and bolsters its efficiency in swiftly and precisely addressing MHD challenges. Additionally, the Titov-Démoulin flux rope model \citep{Titov14} and the force-free flux rope construction using the Regularized Biot-Savart law (RBSL) method \citep{Titov18} in COCONUT accurately describe the initial stages of a CME evolution in the solar corona until O.1 AU \citep{Linan23,Guo24}.

In addition to solar corona simulations, a range of 3D MHD models is dedicated to studying the heliosphere from 0.1 AU to the Earth and even beyond 2 AU. 3D MHD solvers such as ENLIL \citep{Odstrcil03}, MS-FLUKSS \citep{Singh18}, ICARUS \citep{Verbeke22, Baratashvili22},SUSANOO-CME \citep{Shiota16}, LFM-helio \citep{Merkin11,Pahud12} which is time-dependent coupling with the coronal MAS model \citep{Merkin16}, belong to this class of models. The present study focuses on the European Heliospheric Forecasting Information Asset \citep[EUHFORIA;][]{Pomoell18}. This space weather forecasting tool tracks the properties of one or more 3D CME models in the heliosphere after being injected into the domain at 0.1 AU.

The cone model, describing the CME as unmagnetized plasma with a self-similar expansion, was the first CME model implemented in EUHFORIA \citep{Xie04}. The most often used model is the spheromak, representing a force-free magnetic field with a spherical geometry \citep{Chandrasekhar57,Verbeke19}. However, due to the lack of legs, spheromak falls short in accurately representing scenarios where Earth is affected by the flanks of ICMEs. To address this limitation, the "Flux Rope in 3D" \citep[FRi3D;][]{Isavnin16} was implemented by \citet{Maharana22}. This model represents a toroidal flux rope still anchored on the solar surface. Finally, \citet{Linan24} implemented two new toroidal CMEs: one is based on the modified Miller-Turner (mMT) solution \citep{Romashets03,Vandas15} while the other one is derived from the Soloviev equilibrium \citep{Soloviev75}. The latter two models require considerably fewer computational resources than FRi3D, drastically reducing the overall simulation time.

The accuracy of predictions made by heliospheric simulation tools such as EUHFORIA is significantly influenced by two factors: 1) the coronal model used to initiate the solar wind and 2) the characteristics of the CME model employed to simulate the actual events. For the first point, recent findings by \citet{Baratashvili24} demonstrate that using COCONUT as the coronal model instead of WSA results in a bi-modal solar wind structure that matches in-situ measurements. As for the second point, the closer the CME model aligns with the observational data, the more accurate the predictions are expected to be. However, a major challenge arises when injecting the CME model at 0.1 AU, as it omits the critical physical processes in the lower solar corona. Consequently, CMEs do not experience evolutionary changes within the solar corona before entering the model. Furthermore, tracking CMEs within the currently used WSA model is impossible, as this model relies on empirical relationships and is not a time-dependent 3D simulation.

To account for the crucial interaction between the solar wind and the CME within the solar corona, we introduce the first time-dependent coupling between the coronal model COCONUT and the heliospheric forecasting tool EUHFORIA. Specifically, we perform multiple COCONUT simulations where flux ropes evolve from the solar surface to $R_{b}=21.5~\;R_{\odot}$. At each time step, the outer boundary is saved and used as the inner boundary in EUHFORIA. By doing this, the thermodynamic and magnetic properties of CMEs crossing the boundary are directly impacted by their propagation in the lower corona. In the future, we expect predictions made by the space weather chain COCONUT + EUHFORIA will be more accurate than those using CME models independent of the solar corona, such as the spheromak model.

This work presents the {\bf first} coupling between two simulations for test cases. The application to an actual event is reserved for a future study. The paper is organized as follows: the first section is dedicated to describing COCONUT (cf.\ Sect.~\ref{sec:coconut}), with particular emphasis on the numerical scheme (cf.\ Sect.~\ref{sec:fullmhd}), the implemented flux rope models (cf.\ Sect.~\ref{sec:cmeimplementation}), and the handling of outputs (cf.\ Sect.~\ref{sec:outputprep}). Next, we introduce EUHFORIA (cf.\ Sect.~\ref{sec:EUHFORIA}), highlighting the standard version of the solver in Section~\ref{sec:Vanilla EUHFORIA} as well as the numerical changes made to dynamically adjust the inner boundary based on COCONUT's outputs (cf.\ Sect.~\ref{sec:updated euhforia}). In the subsequent section (cf.\ Sect.~\ref{sec:result}), the coupling is tested through a set of six simulations, each varying based on the properties of the flux ropes evolving in COCONUT (cf.\ Sect.~\ref{sec:testcases}). After detailing the propagation of CMEs in COCONUT (cf.\ Sect.~\ref{sec:propagationincoco}), we illustrate how this information is transferred to EUHFORIA (cf.\ Sect.~\ref{sec:interface}) and how the thermodynamic and magnetic properties continue their evolution in the heliosphere up to Earth (cf.\ Sects.~\ref{sec:propagationEUHFORIA} and ~\ref{sec:Earth}). Finally, Section~\ref{sec:conclusion} concludes with a presentation of the main findings of our works and discusses potential future developments.

\section{COCONUT} \label{sec:coconut}
\subsection{Full MHD coronal model} \label{sec:fullmhd}

The first part of our space weather forecasting chain involves tools that generate the solar corona's magnetic field distribution and thermodynamic parameters. This modelling is achieved through the COCONUT simulation. Unlike some other contemporary solvers, COCONUT employs an implicit second-order finite-volume scheme based on the Computational Object-Oriented Libraries for Fluid Dynamics (COOLFluiD) platform \citep{Lani2005,Kimpe2005, Lani2013}. In steady-state simulations, this solver allows for using Courant-Friedrichs-Lewy (CFL) numbers significantly higher than 1 (up to 1000 or more), resulting in up to 30 times faster than state-of-the-art MHD coronal models that use explicit schemes \citep{Perri22}.

COCONUT's high performance also results from using an unstructured grid, which facilitates better handling of regions near the poles by avoiding the presence of singularities. Multiple spatial resolutions are possible. In this case, we used a sixth-level subdivision of the geodesic polyhedron, which results in 1.5 million prismatic cells \citep{Brchnelova2022a}. The mesh extends radially from the inner boundary at $1~\;R_{\odot}$ to the outer boundary at $25~\;R_{\odot}$. For coupling with EUHFORIA, only the surface at $21.5~\;R_{\odot}$ is required. However, the upper boundary of COCONUT is set slightly further away to move the possible outer-boundary effects away from the coupling location \citep{Brchnelova2022b}.

A potential-field approximation is performed using the radial component of the magnetic field derived from a magnetic map to establish an initial condition for the magnetic field. Various magnetograms, such as those from HMI or GONG, can serve as the inner boundary condition \citep{Perri23}. In every instance, the magnetogram is pre-processed by conducting a spherical harmonics decomposition to smooth the magnetic field. Since it is not yet feasible to use a vector magnetogram as the inner boundary condition in COCONUT, which would provide access to the transverse components of the magnetic field, we apply Neumann (zero gradients) conditions across the inner boundary for the colatitudinal and longitudinal magnetic field components and let them evolve freely with the velocity components in such a way, that the resulting velocity is parallel to the background magnetic field \cite{Brchnelova2022b}. The pressure and density are assumed to be constant everywhere \citep{Perri22}.


After initialization, the following conservative formulation of the MHD equations is solved:
\begin{multline} \label{eq:MHDCOCONUT}
\frac{\partial}{\partial t}\left(\begin{array}{c}
\rho \\
\rho \vec{V} \\
\vec{B} \\
E \\
\phi
\end{array}\right)+\vec{\nabla} \cdot \left(\begin{array}{c}
\rho \vec{V} \\
\rho \vec{V} \vec{V}+\tens I\left(P+\frac{1}{2}|\vec{B}|^{2}\right)-\vec{B} \vec{B} \\
\vec{V} \vec{B}-\vec{B} \vec{V}+\underline{\tens I \phi} \\
\left(E+P+\frac{1}{2}|\vec{B}|^{2}\right) \vec{V}-\vec{B}(\vec{V} \cdot \vec{B}) \\
V^2_\text{ref}\mathbf{B}
\end{array}\right) \\ =\left(\begin{array}{c}
0 \\
\rho \vec{g}\\
0 \\
\rho \vec{g} \cdot \vec{V} + \mathbf{S} \\
0
\end{array}\right),
\end{multline}

with $\rho$ the density, $E$ the total energy, $P$ the thermal gas pressure, $\vec g(r) = -(G M_\odot/r^2)\, \hat{\vec e}_r$ the gravitational acceleration, $\tens I = \hat{\vec e}_x \otimes \hat{\vec e}_x + \hat{\vec e}_y \otimes \hat{\vec e}_y + \hat{\vec e}_z \otimes \hat{\vec e}_z$ the identity dyadic and \textbf{S} the energy source term. The solution uses a second-order accurate finite volume discretization and the artificial compressibility analogy \citep{chorin1997,Dedner2002}. The different MHD quantities are normalized by the reference values of $\rho_\text{ref} = 1.67 \cdot 10^{-13}$ kg.m$^{-3}$, $B_\text{ref} = 2.2 \cdot 10^{-4}$ T, $l_\text{ref} = 6.9551 \cdot 10^{8}$ m, and the corresponding $V_{ref}$ computed from the values above.

Until now, the vast majority of studies conducted with COCONUT have assumed 
$\textbf{S}=0$, referring to these as polytropic runs. However, recent work by \citet{Baratashvili24} demonstrates that by incorporating empirical source terms to better approximate the physics of the solar corona, the representation of the solar corona with COCONUT aligns more closely with observations than the purely polytropic version. In light of these findings, we have also decided to use the so-called "full MHD" formulation instead of the polytropic version in this study. The "full MHD" version of COCONUT differs from the polytropic version by adding extra terms on the right-hand side of the energy equation given in Eq.~\ref{fullMHD_terms}. When including the complex physics phenomena in the HMD equations, we switch the adiabatic index to $\gamma = 5/3$, which is a realistic polytropic index compared to the artificially reduced one used in the polytropic simulations. In the "full MHD" version, the same inner boundary conditions are used as in the polytropic COCONUT simulations. Furthermore, the MHD equation set is solved uniformly across the entire numerical domain without differentiating between open and closed magnetic field lines.

In detail, the source terms in the energy equation can be described as follows:
\begin{equation} \label{fullMHD_terms}
    S = - \nabla \cdot \mathbf{q} + Q_{rad} + Q_{H},
\end{equation}
where $Q_{H}$ is the coronal heating, $Q_{rad}$ is the radiation loss function and $- \nabla \cdot \mathbf{q}$ is the thermal conduction. According to \citet{Rosner78}, the radiative loss can be written as : 
\begin{equation}
    Q_{rad} = - n_en_p P(T),
\end{equation}
with $n_e$ and $n_p$ are the electron and proton number densities, and $P(T)$ is a cooling profile depending on the temperature \citep{Rosner78}. 
 
The thermal conduction follows two regimes depending on whether the plasma is collisional or not \citep{Mikic99,Hollweg78}. Below $10~\;R_\odot$, the plasma is collisional, and the thermal conduction is described by the standard Spitzer-H$\ddot{\text{a}}$rm thermal conduction flux :
\begin{equation}
\vec{q}_1 = -\kappa_{\parallel} (\hat{\vec{b}} \otimes \hat{\vec{b}}) \cdot \nabla T
\end{equation}
with $\kappa_{||} = 9\times10^{-7} T^\frac{2}{5}$ in cgs units. Beyond $10~\;R_\odot$, in the regime which is assumed to be collisionless, the thermal conduction flux is then given by: 
\begin{equation}
    \vec{q_2} = \alpha n_e k T \vec{v},
\end{equation}
with $\alpha$ a constant defined in \cite{Hollweg78} and $k$ the Boltzmann constant.

Since the solar coronal heating mechanism is still not fully understood in solar physics, the formulation of the heating term $Q_{H}$ remains under investigation. \citet{Baratashvili24}, using the same magnetogram date as in this work, tested three widely-used empirical formulations for the heating term in COCONUT. First, they tried a function that depends only on the distance from the Sun's surface. However, this uniform spherically symmetric heating did not yield a realistic bi-modal solar wind configuration, so we ruled out this option for this article. \citet{Baratashvili24} also considered a more complex heating function approximation depending on the magnetic field strength and described as : 
\begin{equation} \label{heating_lionello}
    Q_H = Q_H^{exp} + Q_H^{QS}+Q_H^{AR},
\end{equation}
\begin{equation}
    Q_H^{exp} = H_0 e^{\frac{-(r-R_\odot)}{\lambda_0}},
\end{equation}
where $H_0 = 4.9128 * 10^{-7}$ erg cm$^{-3}$ s$^{-1}$, $\lambda_0 = 0.7 R_\odot$, $Q_H^{QS}$ a heating term describing the quiet Sun and $Q_H^{AR}$ a term describing the effect of active regions \citep[cf.][for more details]{Lionello09}. Finally, \citet{Baratashvili24} tried a heating function directly depending on the strength of the magnetic field, as suggested by \citet{Downs2010} and \citet{Pevtsov2003} : 
\begin{equation} \label{magnetic_damping}
    Q_{H} = H_0 \cdot |\mathbf{B}| \cdot e^{-\frac{r-R_s}{\lambda}},
\end{equation}
where $H_0 = 4 \cdot 10^{-5}$ erg cm$^{-3}$ s$^{-1}$ G$^{-1}$ and $\lambda=0.7~\;R_\odot$. Using either Eq.~\ref{heating_lionello} or Eq.~\ref{magnetic_damping} results in a realistic bi-modal solar wind in COCONUT, with a coronal magnetic field distribution that is highly similar between both formulations. To settle on the most suitable expression, \citet{Baratashvili24} connected the COCONUT output to the heliosphere to examine how these effects propagate to Earth. Specifically, these authors used COCONUT output in the heliospheric simulation ICARUS to initialize the solar wind. This is similar to what is done with EUHFORIA in the current work, but for a single-time instance and without CME. By comparing profiles at Earth with OMNI 1-min data, \citet{Baratashvili24} found that the modelled heliosphere in ICARUS was more consistent with the OMNI data using the heating introduced in Eq.~\ref{magnetic_damping} compared to the heating function in Eq.~\ref{heating_lionello}. However, the number density remained significantly overestimated. For this reason, we decided to use the heating term described in Eq.~\ref{magnetic_damping} for this study.

\subsection{CME implementation} \label{sec:cmeimplementation}

This study will track the propagation of multiple CME models in EUHFORIA following their evolution in COCONUT. Tracking a CME's propagation within COCONUT involves several steps. The solar wind in which the CME will propagate must be modelled initially. To achieve this, COCONUT is run in its relaxation mode, during which the full MHD equations are solved until a quasi-steady state is reached. The convergence is monitored by assessing the global residuals of various physical quantities according to the following formula:
\begin{equation} \label{eq:residue}
{\rm res}(a)={\rm log}\sqrt{\sum_{i}(a_{i}^{t}-a_{i}^{t+1})^{2}},
\end{equation}
where $a$ represents the considered primitive variable, $i$ is the spatial index, and $t$ is the temporal index.

Once the solar wind is established, the magnetic field is modified to superimpose that of a CME model at the solar surface \citep[cf.][for details on the implementation of a flux rope in COCONUT]{Linan23}. It is important to note that only the magnetic field is altered during the implementation. The density, velocity, and temperature remain unaffected by adding a CME model.

Currently, two different CME models can be implemented in COCONUT. The first model, whose implementation is described in \citet{Linan23}, is the modified Titov-Démoulin model \citep[hereafter, TDm]{Titov14}. It represents a current-carrying and approximately force-free magnetic field along a toroidal geometry. The toroidal segment is partially embedded in the photosphere, with only one circular arc extending into the corona. The magnetic twist is concentrated in a thin layer near the boundary. Moreover, during the implementation, the flux rope's plane is positioned locally perpendicular to the ambient magnetic fields, which should be potential.

Since no initial velocity is imposed, the flux rope must not be initially in equilibrium to allow for radial evolution. This is the case when the tension of the overlying magnetic field does not balance its magnetic pressure. Practically, this condition is met when the net intensity of the flux rope, $I$, is set higher than the Shafranov intensity as follows:
\begin{equation}
  I = \zeta I_{S},
\end{equation}
with 
\begin{equation} \label{eq:IS}
  I_{S}\approx \frac{4\pi RB_{\perp}/\mu}{\ln{\frac{8R}{a}}-\frac{3}{2}+\frac{I_{i}}{2}}.
\end{equation}
Here, $R$ is the major radius of the torus, $a$ is the minor radius, $I_{i}$ represents the internal self-inductance per length of the rope, $\zeta$ is a positive integer, and $B_{\perp}$ is the ambient magnetic field perpendicular to the toroidal axis. When $\zeta>1$, the ambient magnetic field cannot prevent the expansion of the toroidal segment.

Unlike the TDm model, which follows a toroidal geometry, the second CME model allows for a flux rope with an arbitrary path, the RBSL model. This model was originally proposed by \citet{Titov18} and implemented in COCONUT by \citet{Guo24}. The magnetic field can be expressed as:
 \begin{eqnarray}
\boldsymbol{B}=\nabla \times \boldsymbol{A_{I}} + \nabla \times \boldsymbol{A_{F}}, \label{eq7} \\
\boldsymbol{A_{I}(x)}=\frac{\mu I}{4\pi}\int_{C\cup C^{'}}K_{I}(r)\boldsymbol{R^{'}}(l)\frac{dl}{a(l)}, \label{eq8} \\
\boldsymbol{A_{F}(x)}=\frac{F}{4\pi}\int_{C\cup C^{'}}K_{F}(r)\boldsymbol{R^{'}}(l)\times\boldsymbol{r}\frac{dl}{a(l)^2}, \label{eq9}
 \end{eqnarray}
with $a$ the minor radius, $C$ the axis path, $C^{'}$ the mirror path of the axis relative to the photosphere, $l$ the arc length, $\boldsymbol{R}$ the radius-vector, $\boldsymbol{R^{'}}$ the tangential unit vector and $r$ the vector from the source point to the field point. According to \citet{Titov18}, $K_{I}(r)$ and $K_{F}(r)$ are the kernels of the RBSL model, which can be written as piecewise functions: 
\begin{eqnarray}
 K_{I}(r)=  \begin{cases} \frac{2}{\pi}(\frac{\rm{arcsin r\ }}{r}+\frac{5-2r^{2}}{3} \sqrt{1-r^{2}}) \qquad 0 \le r \le 1 \\
 \frac{1}{r} \qquad  r > 1, \end{cases} \label{eq10}
 \end{eqnarray}
and
\begin{eqnarray}
 K_{F}(r)=  \begin{cases} \frac{2}{\pi r^{2}}(\frac{\arcsin r}{r}-\sqrt{1-r^{2}})+\frac{2}{\pi}\sqrt{1-r^{2}}
 + \\
 \frac{5-2r^{2}}{2\sqrt{6}}[1-\frac{2}{\pi} \arcsin(\frac{1+2r^{2}}{5-2r^{2}})] \qquad  0 \le r \le 1,\\ 
 \frac{1}{r^{3}} \qquad  r > 1. \end{cases} \label{eq11}
 \end{eqnarray}
The magnetic field is force-free with this particular formulation, and the net axial current follows a parabolic distribution.
 
As mentioned previously, the path of the RBSL model can be controlled. In this work, this path follows the equations defined in \citet{Torok10} and \citet{Xu20}:
 \begin{eqnarray}
 && f(s)=  \begin{cases} \frac{s(2x_{c}-s)}{x_{c}^{2}}\theta, \quad 0 \le s \le x_{\rm c} \\ \frac{(s-2x_{_{\rm c}}+1)(1-s)}{(1-x_{c})^{2}}\theta, \qquad x_{_{c}} < s \le 1 \end{cases} \label{eq12}
  \end{eqnarray}
with
\begin{eqnarray}
   x&=&(s-x_{c})\cos f + x_{c}, \label{eq13}\\
   y&=&(s-x_{c})\sin f, \label{eq14}
 \end{eqnarray}
 and
 \begin{eqnarray}
 && z(x)=  \begin{cases} \frac{x(2x_{\rm h}-x)}{x_{\rm h}^{2}}h, \quad 0 \le x \le x_{\rm h} \\ \frac{(x-2x_{_{\rm h}}+1)(1-x)}{(1-x_{\rm h})^{2}}h, \qquad x_{_{\rm h}} < x \le 1 \end{cases} \label{eq15}
 \end{eqnarray}
where $x_{c}$ is the intersection point of the projected curve on the COCONUT surface and the line connecting the two footpoints, $\theta$ is the angle defining the orientation of the tangent vector at the position $x_{c}$, $x_{h}$ is the position of the apex, which is at the height $h$. The implementation of the RBSL model in COCONUT is further detailed in \citet{Guo24}.

Once the CME is implemented using the TDm or RBSL model, COCONUT is relaunched in its time-accurate mode \citep{Linan23}. In this mode, the backward Euler scheme used to reach a steady state is replaced by a backward differentiation formula (BDF2) with a time step fixed at $5.10^{-3}$ (in code units). This particular value was chosen by \citet{Linan23} and \citet{Guo24} as it ensures accurate simulation without excessive computational resource usage. The linearized system was solved using the Generalized Minimal RESidual (GMRES) method, complemented with a parallel Additive Schwartz preconditioner from the PETSc library toolkit\footnote{\url{https://petsc.org/release/}}. The convergence of the time-accurate iterative process at each time step was determined either when the residual fell below $1 \times 10^{-4}$ or after ten sub-iterations. These settings were chosen after extensive testing because they allow for a fast and accurate simulation. Using more stringent values increased run-time excessively without significantly improving accuracy. Finally, intermediate simulation results (i.e. solution in terms of primitive variables) are saved every 20 iterations, corresponding to approximately $144$ seconds in physical time.

\subsection{Output pre-processing} \label{sec:outputprep}

\begin{figure*}[htb!]
    \centering
    \includegraphics[width=1\textwidth]{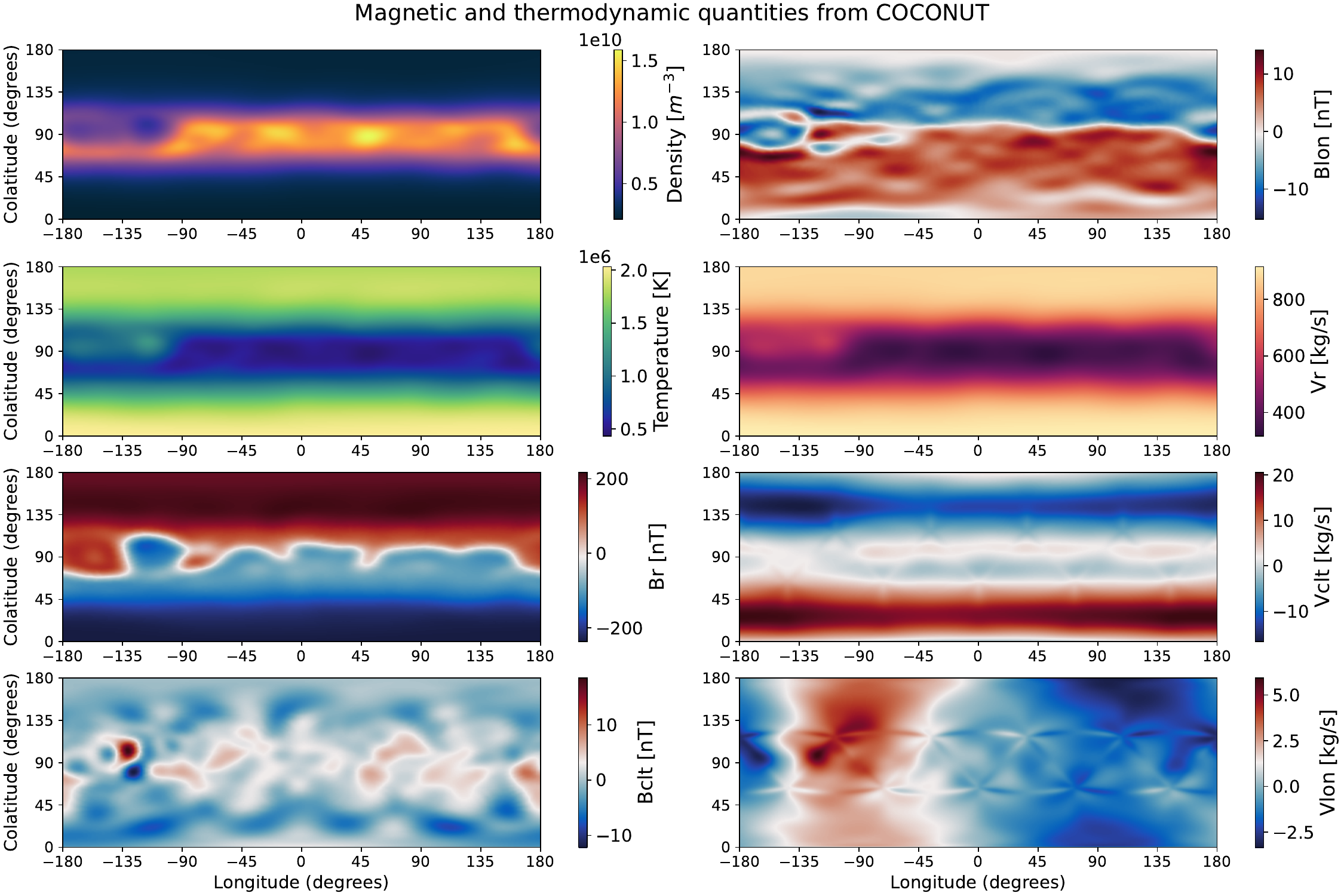}
    \caption{Magnetic and thermodynamic quantities derived from the surface at $R=21.5R_{\odot}$ in the COCONUT simulation. The panels represent the density, the temperature, the three components of the magnetic field ($B_{r}$ $B_{clt}$ and $B_{lon})$, and the three components of the velocity ($V_{r}$ $V_{clt}$ and $V_{lon})$. These panels show the surface at the initial moment of the COCONUT simulation, where a CME modelled by the TDm model with $\zeta=35$ is implemented. These interpolated surfaces, derived from the unstructured mesh of COCONUT and saved in a coupling text file, are used throughout the relaxation phase to build up the solar wind in EUHFORIA.}
    \label{fig:initialtime}
\end{figure*}

Currently, the process of running the COCONUT and EUHFORIA simulations remains sequential: it is necessary first to run COCONUT entirely, followed by EUHFORIA, as integration into a single execution has not yet been implemented. During the execution of COCONUT, the evolution of various thermodynamic quantities must be regularly saved for later use in EUHFORIA. In this paper, we  present for the first time a time-dependent coupling between COCONUT and EUHFORIA.

COCONUT offers the ability to save several types of files. Among these options, we have retrieved the output as VTU files (VTK Unstructured Grid File). These files can be read in the ParaView software \citep{Squillacote07}, allowing us to visually verify, for example, by tracing magnetic field lines, that the simulation is proceeding correctly, which means that the CME implemented develops radially in the domain over time \citep[e.g. Fig. 5 in][]{Linan23}. 

The coupling between the two simulations requires using the native "CFmesh" files generated by COCONUT. In particular, these ASCII files contain the mesh nodes' coordinates, the cell-node connectivity, and the values of the thermodynamic quantities at the cell centres. Unlike the VTU files, a single CFmesh file is generated and written in parallel for the entire domain at each iteration, as the solver directly integrates the information processed by the different processors.

As will be mentioned in Section~\ref{sec:updated euhforia}, EUHFORIA requires the values of the magnetic field, velocity, density, and temperature at the surface that is located precisely at $21.5~\;R_{\odot}$. Since the mesh in COCONUT is unstructured, to extract this specific surface, we proceed to interpolate the unstructured grid into a structured grid corresponding to $R_{b}=21.5~\;R_{\odot}$ with 360 points in longitude and 180 in latitude.

The interpolation is performed upstream of EUHFORIA using the radial basis function interpolation method with a linear kernel. To determine the value at a given point, the interpolation is based on the values of the 50 cells nearest to that point on the original grid.

The interpolated surface, illustrated in Fig.~\ref{fig:initialtime}, is saved in a new ASCII file (hereafter the coupling file or boundary map) that EUHFORIA will read. This step is repeated for all iterations saved by COCONUT, generating many text files representing the surface's temporal evolution at $21.5~\;R_{\odot}$. During the creation of the files, a 180-degree rotation in longitude is applied so that the injected CME is directed towards the Earth in EUHFORIA. It should be noted that this interpolation step is computationally costly. Future work is necessary to optimize the data transfer from COCONUT to EUHFORIA.

We also note that the surfaces associated with longitudinal and colatitudinal speed components, cf.\ the last two panels in Fig.~\ref{fig:initialtime}, exhibit numerical artefacts. These artefacts are created due to mesh non-orthogonalities causing the formation of spurious fluxes in non-dominant dynamics, as discussed in detail in \cite{Brchnelova2022a}. The solar wind is radial at the front boundary before the CME arrives. However, the presence of these artefacts does not affect the results, which will be presented later.

\section{EUHFORIA} \label{sec:EUHFORIA}

The following subsections describe the functionality and numerical implementation of the heliospheric model used in this work, EUHFORIA. First, we will describe the standard version of EUHFORIA (cf.\ Sect.~\ref{sec:Vanilla EUHFORIA}). Then, we will discuss the updated version of EUHFORIA that enables coupling with COCONUT (cf.\ Sect.~\ref{sec:updated euhforia}).

\subsection{standard version of EUHFORIA} \label{sec:Vanilla EUHFORIA}
\subsubsection{Numerical scheme}

COCONUT enables tracking the evolution of CMEs in the solar corona up to $21.5~\;R_{\odot}$. To extend the study beyond this limit and predict the temporal evolution of various plasma quantities at different positions in the inner heliosphere, we couple COCONUT with the physics-based 3D MHD heliosphere model, EUHFORIA \citep{Pomoell18}.

The interface between the two simulations was set at $21.5~\;R_{\odot}$ because at this distance, we expect a supersonic and super-Alfvénic solar wind, meaning that no information travels towards the Sun. However, for a physical coupling, it is crucial to ensure that the solar wind is not sub-Alfvénic in COCONUT.

In EUHFORIA, the thermodynamic and magnetic quantities are self-consistently computed by solving the set of ideal MHD equations with gravity. The equations solved are similar to those presented for the COCONUT solver, with the critical difference that EUHFORIA is purely polytropic and does not solve the last equation in Eq.~\ref{eq:MHDCOCONUT}, which corresponds to the conservation equation for magnetic flux for hyperbolic divergence cleaning. In other words, no source term is explicitly implemented ($S=0$). To model the heating that leads to the acceleration of the solar wind, the polytropic index appearing in the definition of the energy, $E = \frac{P}{(\gamma - 1)}P + \frac{1}{2} \rho v^2 + \frac{B^2}{2 \mu_0}$ is taken as a non-adiabatic polytropic index \citep[i.e $<5/3$;][]{Pneuman71,Pomoell12}. As in \citet{Odstrcil04}, the reduced index is set to 1.5.

The MHD equations are solved in the Heliocentric Earth Equatorial (HEEQ) coordinate system using a finite-volume method combined with a constrained transport approach to ensure that the magnetic field remains divergence-free. To achieve a robust second-order accurate scheme, standard piece-wise linear reconstruction and an approximate Riemann solver are employed \citep{Kissmann12}. The Coriolis and centrifugal effects of using a non-inertial frame are neglected, assuming their contributions are negligible \citep{Pomoell18}.

The mesh used in this work is uniform and extends from 0.1 AU (i.e. $21.5~\;R_{\odot}$) to 2 AU with 256 cells in the radial direction and a $2^\circ$ angular resolution. While the computational domain spans $360^\circ$ in longitude, it covers only latitudes from $-60^\circ$ to $-60^\circ$ to avoid polar regions.

\subsubsection{Boundary conditions} \label{sec:boundary}

To establish the background solar wind in the heliosphere, the standard version of EUHFORIA reads the surface at $21.5~\;R_{\odot}$ which is stored in a single boundary condition map after being generated by a coronal model. The standard EUHFORIA as described by \citet{Pomoell18}, only requires the 2D distribution of the radial magnetic field, $B_{r}$, radial velocity $V_{r}$, density, $n$, and temperature, $T$ as inner boundary condition. In most of the studies using EUHFORIA \citep[e.g.][]{Maharana23,Verbeke19,Prete2024,Linan24}, the semi-empirical WSA  model \citep{McGregor11,van10} is used to establish the boundary condition map. This work aims to demonstrate how the solar corona simulation COCONUT can be used in place of the semi-empirical WSA model within EUHFORIA. However, this subsection explains how EUHFORIA currently uses the WSA boundary condition map to provide context for understanding the modifications made to the code to read the series of boundary maps generated by COCONUT (cf.\ Sect.~\ref{sec:updated euhforia}). 

In the WSA model, a Potential Field Source Surface (PFSS) extrapolation \citep{Altschuler69} is performed using the magnetic field derived from a synoptic magnetogram. The magnetic field obtained is then extended up to 0.1 AU following the Schatten Current Sheet model \citep{Schatten69}. The solar wind speed $v_{sw}=v(f,d)$ is determined based on the flux tube expansion factor $f$ and the distance $d$ from the footpoint of the flux tube to the nearest coronal hole. Using this solar wind speed, the density and temperature at the outer boundary are calculated through empirical relations. These plasma quantities (i.e. $B_{r}$, $V_{r}$, $n$ and $T$) are then saved in an inner boundary file \citep{Pomoell18}. This step is performed upstream of the main EUHFORIA run. The single file thus created is used at each time step in the standard EUHFORIA as the inner boundary condition.

At the inner boundary of EUHFORIA, two layers of ghost cells are used. The values of these ghost cells are determined at each step through linear extrapolation using the boundary condition map and the data from the first in-domain cell. In addition to the information contained in the boundary file when the WSA model is used as a coronal model in EUHFORIA, we assume that the velocity is purely radial at the interface $R_{b}=21.5~;R_{\odot}$. Thus, the colatitudinal velocity component, $V_{clt}$ and longitudinal velocity, $V_{lon}$, are set to zero. The same assumption is made for the colatitudinal magnetic field component. The longitudinal magnetic field is determined using the radial magnetic field and the radial velocity according to the Parker Solar Wind model \citep{Parker58}:

\begin{equation} \label{eq:Bl}
    B_{lon}=R_{b}\frac{\Omega B_{r}}{V_{r}}\sin{\theta},
\end{equation}
where $\theta$ is the colatitude, and $\Omega$ is the solar rotation rate, approximately equal to $2.66622373 \times 10^{-6} \, \text{rad} \, \text{s}^{-1}$.

Open boundary conditions are applied at the outer radial boundary by extrapolating the existing values beyond the domain's edge. In contrast, a symmetric reflection of values is employed at the latitudinal boundaries to ensure coherent continuity across these boundaries.

In the standard version of EUHFORIA, which uses the semi-empirical WSA coronal model, the same boundary condition file is used throughout the simulation. To model the temporal evolution of the solar corona, a simple rotation of the boundary map is applied, assuming that the coronal model remains static during the run.

\subsubsection{Implementation of CME models}

In the standard version of EUHFORIA, the initial time, $t=0$, corresponds to the observation time of the magnetogram used to create the boundary map via a coronal model. When $t$ is positive, the simulation enters the forecast phase to predict the evolution of one or more CMEs in the heliosphere. However, the solar wind must be already established at $t=0$ before tracking CMEs within the domain. Thus, the forecast phase is preceded by a relaxation phase to establish a stable solar wind solution that conforms to the prescribed inner boundary conditions. This phase allows the solar wind to evolve within the simulation until it reaches a steady-state configuration, ensuring the model starts from a consistent and realistic background. During this time, the simulation iterates over several steps, adjusting the plasma properties until the solution no longer changes significantly, creating a balanced initial condition for the heliospheric plasma. Typically, this process takes around 14 days of simulation time, sufficient for a solar wind moving at approximately 250 km/s to travel through 2 AU and reach the outer edge of the simulation domain \citep{Pomoell18}.

During this phase, when EUHFORIA used the WSA model as coronal model the single boundary map is rigidly rotated at each time step by an angle equal to $t\Omega$ ($t$ being negative) to maintain the domain in the HEEQ coordinate system. Once the solar wind is established and before the forecast phase, it is possible to insert into the domain all the CMEs that have occurred before the start of the forecast. Since the passage of a CME affects the heliospheric plasma for several days, this ensures that the initial conditions accurately reflect recent perturbations in the spatial environment. By integrating these CMEs, the accuracy of the forecasts is improved by accounting for significant changes in the plasma and magnetic fields caused by these events.

Presently, five different CME models are implemented in EUHFORIA, each with distinct features and limitations. The simplest is the cone model \citep{Xie04}, which represents an unmagnetized plasma with self-similar expansion, useful for modelling the interplanetary acceleration of low-energy protons during SEP events but unsuitable for studying internal magnetic signatures of ICMEs \citep{Wijsen22}. The spheromak model depicts a CME with a global spherical shape and a linear force-free magnetic field \citep{Chandrasekhar57,Verbeke19}, widely used but inadequate for scenarios where the flanks or legs of ICMEs impact Earth. The flux rope model \citep[FRi3D;][]{Maharana22,Isavnin16} features an extended flux-rope geometry, including CME legs, closely matching observational data of crescent-shaped CMEs. However, it requires substantial computing time, limiting its operational use. Additionally, two toroidal CME models have been implemented to address the limitations of previous models by incorporating toroidal structures, enhancing accuracy and computational efficiency in representing CME impacts \citep{Linan24}.

Regardless of the chosen CME model, it is inserted into EUHFORIA by making it pass through the inner boundary at $R_{b}=21.5~\;R_{0}$. A function mask identifies the grid points where the CME intersects the inner surface. At these intersection points, the solar wind speed, magnetic field, density, and temperature are replaced by those of the CME. At each time step, the CME centre advances with the initially defined radial speed, updating the 3D mask region as the CME moves through the boundary. Injecting CME models in this manner can potentially lead to an added divergence in the magnetic field.

\subsection{Updated EUHFORIA} \label{sec:updated euhforia}

\begin{figure}[ht!]
    \centering
    \includegraphics[width=0.5\textwidth]{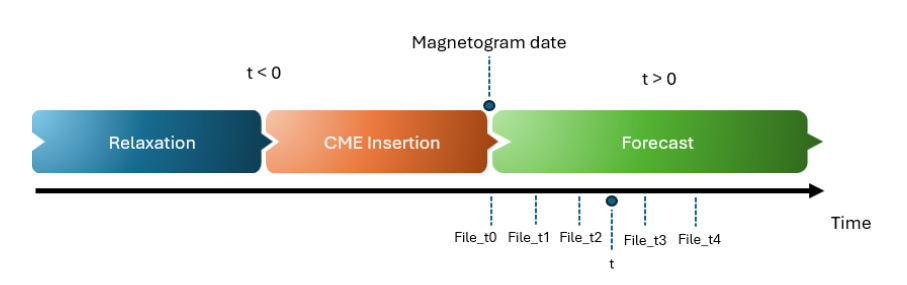}
    \caption{The different phases of an updated EUHFORIA run. The same inner boundary file is used to construct the solar wind throughout the relaxation phase. From the magnetogram date corresponding to the first coupling file, the inner boundary ghost cells of EUHFORIA at a given time $t$ are updated using temporal interpolation of the information in the two files bracketing this particular time $t$.}
    \label{fig:runphase}
\end{figure}

In this subsection, we will describe the code changes to EUHFORIA to enable coupling with the evolving coronal simulation, COCONUT. Unlike a static coronal model such as WSA, the thermodynamic and magnetic properties at the boundary $R_{b}=21.5R_{\odot}$ in COCONUT can evolve, reflecting the arrival and passage of a CME that is introduced at the Sun’s surface (cf.\ Sect.\ref{sec:cmeimplementation}). As a result, a single file can no longer serve as the inner boundary condition for the entire EUHFORIA run, and the code has been adapted to handle a series of files, each containing the $R_{b}=21.5~R_{\odot}$ surface of COCONUT at specific time steps. These files will be referred to as $file_{t0}$, $file_{t1}$, and so on, where $file_{t0}$ contains the information for the 2D surface at the initial time of the magnetogram, and $file_{t1}$ contains the information for the time $t_0+1\times dt$, and so forth.

The new version of EUHFORIA, like the standard version, is still divided into three main phases: the relaxation phase, the CME insertion phase, and the forecast phase (cf.\ Fig.~\ref{fig:runphase}).

As previously described (cf.\ Sect.~\ref{sec:boundary}), EUHFORIA begins with a relaxation phase to establish the solar wind state across the entire domain at the initial time $t=0$, based on information provided at the inner boundary and read from a boundary map. Since the first coupling file, $file_{t0}$ contains the magnetic and thermodynamic information at the interface between COCONUT and EUHFORIA at the initial time, this file is used throughout the relaxation phase to initialize the simulation.

In the standard version of EUHFORIA using the semi-empirical WSA model, it is assumed that at the inner boundary of EUHFORIA, the velocity is purely radial, the longitudinal magnetic field is proportional to the radial magnetic field (cf.\ Eq.~\ref{eq:Bl}), and the colatitudinal magnetic field is zero (cf.\ Sect.~\ref{sec:boundary}). However, COCONUT provides all velocity and magnetic field components, which are crucial for accurately describing the propagation of a CME. As a result, in the updated version of EUHFORIA that uses COCONUT as the coronal model, no assumptions are made about the magnetic field and velocity components. This information comes directly from COCONUT. The three components of the magnetic field, the three velocity components (cf.\ Fig.~\ref{fig:initialtime}), and temperature and density are stored in the coupling files. Temperature is not a primitive variable in COCONUT; it is computed using the ideal gas law for pressure and density. The EUHFORIA code was modified to read these eight variables to update the inner boundary instead of the previous four (cf.\ Sect.~\ref{sec:boundary}).

Starting from time $t=0$, which corresponds to the magnetogram date and the end of the relaxation phase, EUHFORIA enters its forecast phase and updates the inner boundary at each time step using the information from the series of coupling files. However, EUHFORIA’s time step is not fixed; it varies to ensure numerical stability, as it is adjusted according to the CFL condition, which depends on the speed of disturbances within the domain—faster propagation requires smaller time steps. Since the exact time steps of the simulation are not known in advance, it is challenging to have a coupling file, $file_{ti}$, corresponding precisely to every simulation time $t$. Therefore, the code identifies the two files that bracket the current simulation time and performs a linear interpolation of the data in these two boundary maps to determine the required quantities. Having as many coupling files as possible, with minimal temporal gaps between them, is recommended to ensure interpolation accuracy.

The inner boundary of EUHFORIA continues to update at each time step until the simulation time exceeds the time stored in the last coupling file. At that point, and for the remainder of the forecast phase, the last available boundary map is used as the inner boundary condition, similar to how the first coupling file was used during the relaxation phase. A rigid rotation of this final coupling file is performed at each time step to model the temporal evolution of the inner boundary.

Finally, it is important to note that inserting CME models (e.g., spheromak) as done in the standard version of EUHFORIA is still possible. However, additional studies are needed to determine how these models might interact with the CMEs propagated in COCONUT and transmitted to EUHFORIA via the boundary maps.

\section{CME propagation from Sun to Earth} \label{sec:result}

\subsection{Test cases} \label{sec:testcases}

\begin{figure}[h!]
    \centering
    \includegraphics[width=0.5\textwidth]{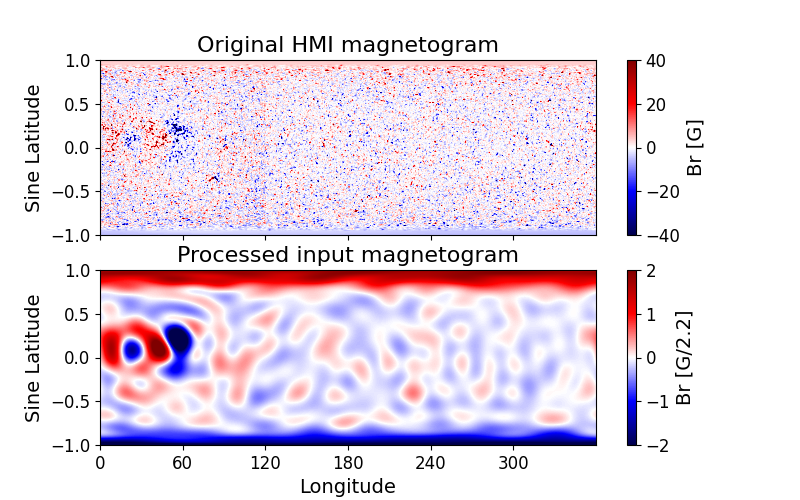}
    \caption{Synoptic magnetogram for July 2, 2019. The upper panel corresponds to the original filled HMI magnetogram, while the bottom panel shows the processed magnetogram used as input in COCONUT.}
    \label{fig:HMI}
\end{figure}

To demonstrate the coupling of COCONUT and EUHFORIA, we study the propagation of different CMEs from the Sun to the Earth. We focus on verifying that the information in COCONUT is accurately transmitted to EUHFORIA.

The magnetic field in COCONUT before implementing the CME is obtained via a steady-state run based on the pole-interpolated Helioseismic and Magnetic Imager (HMI) magnetogram product for July 2, 2019. To be used as input to COCONUT, the HMI magnetogram has been processed with spherical harmonics decomposition filtered with $l_{max} = 20$ \citep[see][for more details about the choice of this particular value and how the projection works.]{Kuźma23}. The pre-processing smooths and homogenizes the magnetic field. However, significant negative and positive polarities exist between longitudes 0 and 60 (cf.\ Fig.~\ref{fig:HMI}).

This particular date was chosen because \citet{Baratashvili24} demonstrate, by comparing with white light images, that the full MHD version of COCONUT effectively reproduces the solar coronal magnetic field distribution. Additionally, 2019 corresponds to a minimum of solar activity, which helps minimize the effects of the background solar wind on CME propagation. Moreover, the simplicity of the magnetic configuration during the solar minimum enhances numerical convergence.\citet{Baratashvili24} also showed that the output from COCONUT for this particular date could be used to create the solar wind for a heliospheric simulation, ICARUS. In the future, the coupling should also be tested in a maximum solar activity. However, during such intense activity, we expect significant interactions between the solar wind and flux ropes \citep{Linan23}, adding unnecessary complexity for an initial introduction to the COCONUT-EUHFORIA coupling, as we present here.

\begin{figure}[h!]
    \centering
    \includegraphics[width=0.5\textwidth]{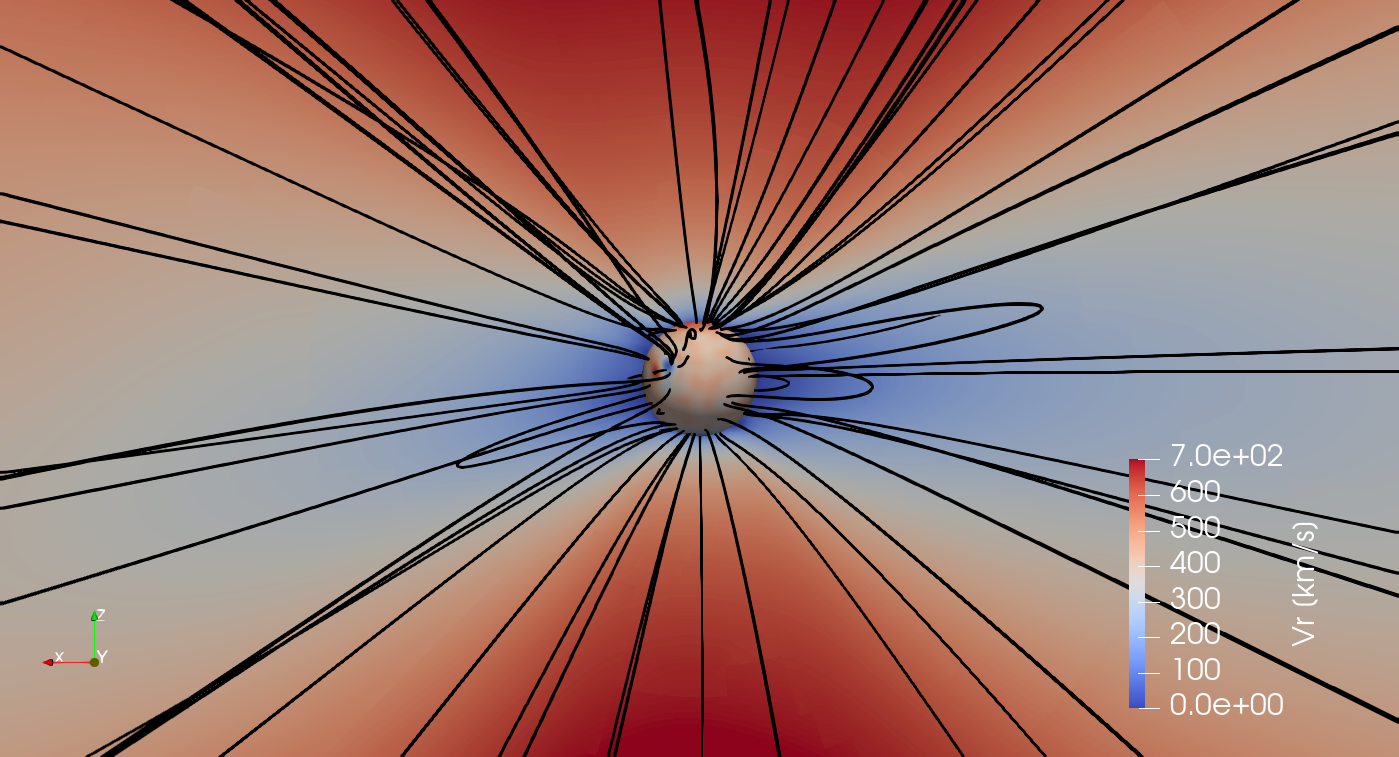}
    \caption{Meridional plane of the radial velocity after the steady-state run of the full MHD COCONUT simulation. The black lines represent a sample of the magnetic field lines of the solar wind.}
    \label{fig:solarwind}
\end{figure}

Figure~\ref{fig:solarwind} shows the radial magnetic field in the meridional plane obtained after the steady-state run for the July 2, 2019, magnetogram. The distribution is bi-modal because the speed is faster at the poles than near the equator. This is due to more intense heating at the poles, where the magnetic field is generally stronger \citep[cf.][for more details about the impact of heating on the solar wind for this particular date]{Baratashvili24}.

\begin{figure}[t!]
    \centering
    \includegraphics[width=0.5\textwidth]{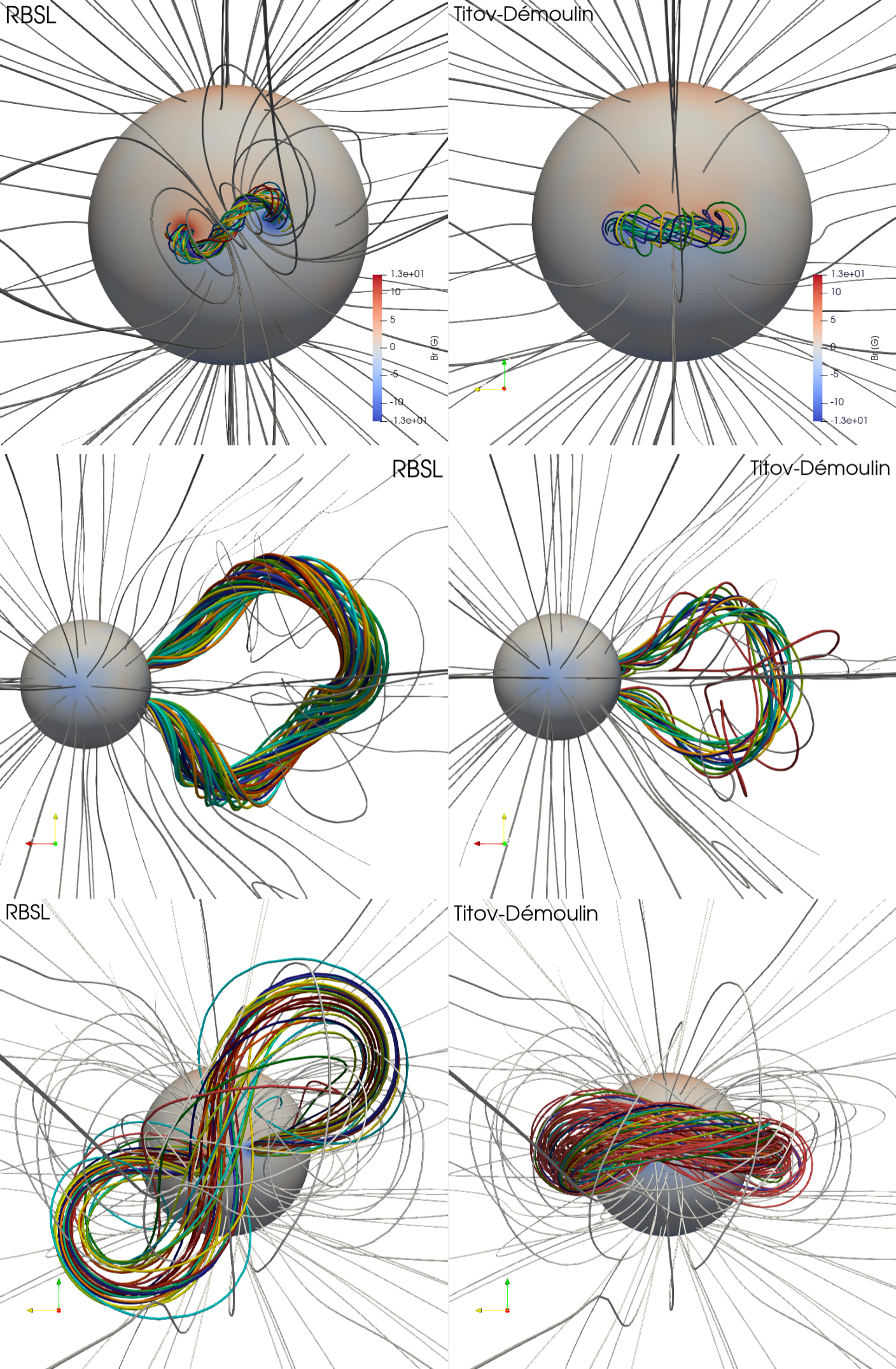}
    \caption{Visualization of the TDm and RBSL models implemented in COCONUT. The upper panels show the CME models just after 
    their insertion in the solar wind, while the other panels show their expansion 52.8 minutes after the simulation starts in the top and side views. The coloured lines represent a sample of magnetic field lines from the flux ropes. The origin for tracing these field lines is a sphere with a radius of $R=0.1R_{\odot}$, positioned at the positive polarity. The grey field lines represent the magnetic field from the background solar wind. The radial magnetic field at the solar surface (measured in Gauss) is also displayed.}
    \label{fig:titovrbsl}
\end{figure}

In the background magnetic field, we implement the two different CME models available: the Titov-Démoulin flux rope model and the RBSL model (cf.\ upper panels in Fig.~\ref{fig:titovrbsl}). We aim to demonstrate that the coupling works regardless of the CME model. In detail, we conducted six simulations, three with each of the two CME models (cf.\ Tab.~\ref{tab:case}). In all cases, the CME is positioned at a longitude of $-180^\circ$ and a latitude of $0 ^\circ$ to ensure evolution mainly in the equatorial plane. This choice of longitude was made to avoid any possible disturbances caused by a high-speed stream located on the opposite side of the Sun at that time (cf.\ Sect.~\ref{sec:propagationincoco}).

\begin{table}[h!]
\centering
\begin{tabular}{cccc}
 & \multicolumn{3}{c}{Initial parameters} \\ \hline
Name &  $\zeta$ &  {$F$ $[10^{20} Mx]$} & $V_{0}$ $[km/s]$ \\ \hline
TDm\_1 &  {35} &  {5.6} &  {1138}   \\ 
TDm\_2 &  {50} &  {8.0} &  {1453}   \\ 
TDm\_3 &  {70} &  {11.2} &  {1780}    \\ 
RBSL\_1 &  {/} &  {8} &  {1263}   \\ 
RBSL\_2 &  {/} &  {12} &   {1580}   \\ 
RBSL\_3 &  {/} &  {15} &  {1804}   \\ 
\end{tabular}
\caption{This table, summarizing the different simulations in our study, shows the parameter $\zeta$ for the TDm cases, the initial magnetic flux $F$, and an approximation of the initial velocity $V_{0}$. The initial velocity was extracted during the first saved time step (i.e., $t=2.412$ min).}
\label{tab:case}
\end{table}

As in \citet{Linan23}, the flux ropes modelled using the TDm formalism have a major radius of $R=0.3~\;R_{\odot}$, a minor radius of $a=0.1~\;R_{\odot}$, and their centre is offset by $d=0.15~\;R_{\odot}$ from the solar surface. Implementing the model results in the addition of two polarities with an area equal to $4839\;$Mm$^{2}$ on the solar surface (cf.\ upper right panel in Fig.~\ref{fig:titovrbsl}). The three CMEs implemented using the TDm flux rope model differ only by the parameter $\zeta$ used, which takes values of $35$, $50$, and $70$. Increasing the $\zeta$ parameter increases the ratio between the magnetic field of the flux rope and the ambient magnetic field. Thus, as described in Section~\ref{sec:cmeimplementation}, setting $\zeta>1$ ensures that the tension of the overlying magnetic field does not prevent the expansion of the flux rope. The higher the $\zeta$ parameter, the higher the magnetic flux. According to \citet{Linan23} and \citet{Regnault23}, the magnetic flux is directly related to the plane-of-sky speed value measured in the simulation's initial moments. This relationship will be confirmed in the following section (cf.\ Sect.~\ref{sec:propagationincoco}). 

For the three implemented RBSL flux ropes, we set $x_{c}=x_{h}=0.5$. The apex of the flux rope is located at a height of $h=0.17~\;R_{\odot}$. As with the TDm CME models, the minor radius is $a=0.1~\;R_{\odot}$. The writhe angle $\theta$ is $60^\circ$. This results in flux ropes with an S-shaped morphology (cf.\ upper left panel in Fig.~\ref{fig:titovrbsl}). In the solar corona, filaments with positive helicity and this particular morphology are frequently observed \citep{Cheng17}.

The parameter $\zeta$ does not appear in the implementation of the RBSL model. Stability is controlled by explicitly modifying the initial magnetic flux. In our work, the three CMEs modelled with the RBSL model have fluxes of $8\times10^{20}~$Mx, $12\times10^{20}~$Mx and $18\times10^{20}~$MX, respectively. Our tests revealed that such fluxes, in this particular solar wind configuration, allow for the expansion of the flux ropes. 

\subsection{Initial stage of CME propagation in COCONUT} \label{sec:propagationincoco}

We initiate the time-dependent COCONUT run once the CMEs are implemented in the steady-state wind. Since the computational resources required for a time-dependent run of COCONUT are significant, we stop the simulation once the magnetic cloud has finished crossing the boundary and there are no significant variations in the different quantities within the domain. Consequently, the simulations stop after $12000$ iterations, corresponding to a physical time span of approximately $24.12$ hours as the CME has fully exited the domain by this point. The 2D surface at $R_{b}=21.5~\;R_{\odot}$ is saved every $20$ iterations, resulting in the generation of $600$ coupling files. 

The two lower panels of Figure~\ref{fig:titovrbsl} show the magnetic field topology after about $t=24.12$ minutes in the COCONUT simulations with the CMEs named "TDM\_3" and "RBSL\_3" (cf.\ Tab.~\ref{tab:case}). The other cases are not represented as they do not show differences in the evolution of the magnetic structure. In Figure~\ref{fig:titovrbsl}, we observe that the evolution is very similar to that described in \citet{Linan23} and \citet{Guo24}, who studied the propagation of the two flux rope models in detail in the polytropic version of COCONUT. The self-consistent radial expansion of the flux rope creates a pinching of the legs, as described in the 2D "standard model" for CME originally developed by \citet{Carmichael64, Sturrock66, Hirayama74, Kopp76} and later extended to 3D by \citet{Aulanier12, Aulanier13, Janvier13}. The current sheet layer formed by the convergence of upward and downward field lines is a region conducive to magnetic reconnection. However, in COCONUT, magnetic reconnection is purely due to numerical dissipation since there is no prescribed resistivity in the set of MHD equations which are solved. Finally, reconnection leads to the formation of post-flare loops, as seen in observations \citep{Schmieder95}.

During the propagation, the TDM flux rope remains in the equatorial plane, while the "S-shape" of the RBSL is preserved and found higher in the solar corona (cf.\ bottom left panel in Fig.~\ref{fig:titovrbsl}). This result aligns with the study by \citet{Guo24b}, which focuses on the impact of flux rope morphology on propagation in COCONUT. This indicates that the magnetic distribution of an ICME is directly linked to its progenitor in solar source regions.

\begin{figure}[h!]
    \centering
    \includegraphics[width=0.5\textwidth]{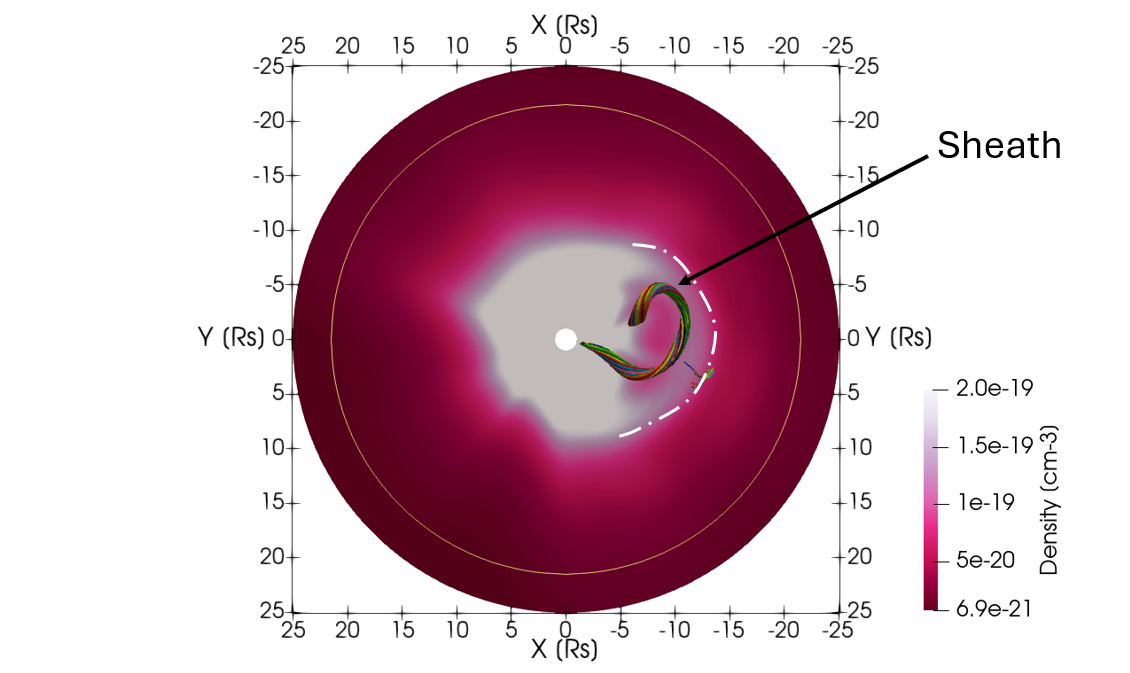}
    \caption{Cross-sections along the equatorial plane of the density for the COCONUT simulation with the TDm flux rope named "TDM\_3". The coloured lines represent a sample of the magnetic field lines of the flux rope. The yellow line indicates the $R_{b}=21.5~\;R_{\odot}$ boundary. A sheath characterized by high density develops ahead of the flux rope during propagation.}
    \label{fig:sheath}
\end{figure}

Among the features shared with those described in the works using the polytropic version of COCONUT \citep{Linan23, Guo24, Guo24b}, we observe an accumulation of matter ahead of the flux rope that intensifies as the CME expands (cf.\ Fig.~\ref{fig:sheath}). This region of heated and compressed solar matter results from the flux rope travelling without allowing enough time for the solar wind to flow around it. This phenomenon is observed by in-situ spacecraft when the CME is sufficiently faster than the surrounding solar wind \citep{Regnault20}.

\begin{figure*}[t!]
    \centering
    \includegraphics[width=\textwidth]{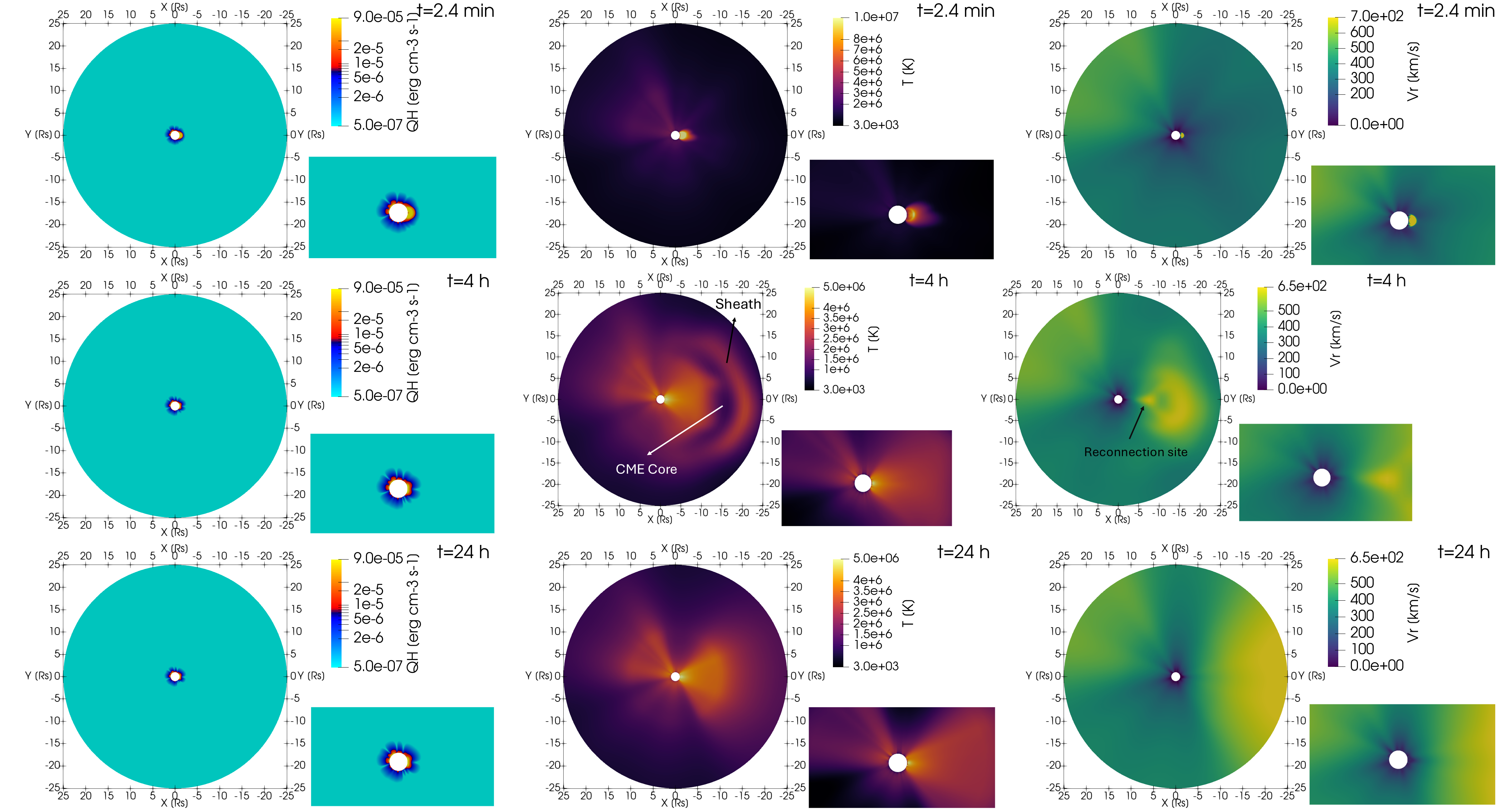}
    \caption{Equatorial plane of the distribution of the heating term, temperature, and radial velocity at different times for the CME named "TDm\_3". The left panels show the heating $Q_{H}$ in  erg cm$^{-3}$ s$^{-1}$ G$^{-1}$ as defined by Equation \ref{magnetic_damping}. The middle panels display the temperature in Kelvin, while the right panels present the radial velocity in km s$^{-1}$. Each panel includes an overall view of the domain and a zoomed-in view near the Sun's surface. From top to bottom, the times represented are $t=2.4$ minutes, $t=4$ hours and $t=24$ hours.}
    \label{fig:heatingdistribution}
\end{figure*}

Figure~\ref{fig:heatingdistribution} presents the equatorial distribution of the heating term, $Q_{H}$, the temperature, and the radial velocity at various times for the simulation with the CME named "TDm\_3". The other cases have common patterns. Shortly after the simulation begins, at the moment $t=2.4$ minutes (i.e., the first saved time step), the tension of the surrounding magnetic field is insufficient to prevent the expansion of the flux rope. This results in a high radial velocity at the flux rope location (cf.\ the top right panel in Fig.~\ref{fig:heatingdistribution}). The initial radial velocities for different simulations can be estimated using this plane. These values are summarized in Tab.~\ref{tab:case}. For simulations involving a TDm flux rope, the velocity ranges from 1138 km/s for the CME with the weakest magnetic flux (TDm\_1) to 1780 km/s for that with the strongest flux (TDm\_3). Similarly, for simulations with an RBSL flux rope, the radial velocity varies from 1263 km/s (TDm\_2) to 1804 km/s (TDm\_3). As noted by \citet{Linan23} and \citet{Guo24}, the greater the magnetic flux, the higher the initial velocity. However, for approximately the same flux, $8\times10^{20} Mx$, the "TDm\_2" case exhibits a lower initial velocity than the "RBSL\_2" case. Since the two CMEs do not share the same geometry, the initial resistance created by the surrounding magnetic field is not exactly the same \citep{Guo2024b}, which could explain this difference.

In the processed magnetogram used as input for COCONUT, we observe strong active regions between $O^\circ$ and $60^\circ$ longitude (cf.\ Fig.~\ref{fig:heatingdistribution}). These regions, characterized by strong magnetic fields, lead to significant heating, as the heating term $Q_{H}$ scales with magnetic field strength, as described in Section~\ref{sec:fullmhd}. Consequently, in the second quadrant (i.e., the region with $X<0$ and $Y>0$) of Figure~\ref{fig:heatingdistribution}, we observe high values of $Q_{H}$. This heating causes particle acceleration, driving high-speed flows, as reported by \citet{Baratashvili24}, and is visible in the last column of the first row in Figure~\ref{fig:heatingdistribution}.

Additionally, at the initial stage, the magnetic field of the flux rope we implemented on the solar surface is stronger than the surrounding magnetic field. In the first row of Figure~\ref{fig:heatingdistribution}, we can logically observe that the heating term, and therefore the temperature, is elevated in the region near the core of the CME.

Later in the simulation, at $t=4$ hours, the CME still resides within the domain. Several temperature zones can be identified (cf.\ middle panel in Fig.~\ref{fig:heatingdistribution}). To the right, near $X=-20~\;R_{\odot}$, a crescent-shaped area of high temperature is visible, which can be linked to the sheath where temperature rises due to the compression of matter ahead of the flux rope's progression. Closer to the Sun, around $X=-15~\;R_{\odot}$, temperatures decrease near the flux rope's core. 

During the implementation of the flux rope in COCONUT, we added two opposite polarities on the surface of the Sun. As a result, at $t=4$, near the initial insertion region, temperatures remain elevated due to coronal heating at the CME footpoints.

It is also noted that the heating associated with $Q_{H}$ remains localized near the surface, as it decreases exponentially with increasing radial distance $r$ (cf.\ Eq.~\ref{magnetic_damping}).

Regarding the velocity distribution, two high-speed regions are observed (cf.\ middle right panel in Fig.~\ref{fig:heatingdistribution}). The first is located at the bulk of the CME, which continues to expand, while the second is localized in the wake of the CME. This rapid plasma flow is associated with magnetic reconnection occurring where opposite magnetic field lines forming the legs of the CME converge \citep{Ando15}.

By the last moment of the simulation at $t=24$ hours, it is assumed (cf.\ Sect.~\ref{sec:propagationEUHFORIA}) that the CME has already traversed the domain. Despite this, high-speed streams created during the CME's wake are observed (cf.\ bottom right panel in Fig.~\ref{fig:heatingdistribution}). The simulation was continued for an additional 24 hours with no noticeable changes in the velocity distribution. This is due to a major limitation of the current COCONUT version, where the solar surface's magnetic field remains fixed throughout the simulation. Consequently, the footpoints of the CME remain on the solar surface even though the flux rope is no longer anchored to these polarities \citep[cf.][for more details]{Linan23}. Enhanced magnetic fields in the active region initially implemented lead to significant solar coronal heating. As a result, due to plasma acceleration, the final distribution features two high-speed streams: one in the second quadrant, caused by the active regions initially present on the solar surface as explained before, and the other in the X-direction, generated by the constant heating of the solar corona at the initial CME footpoints, which does not dissipate during the simulation.

\subsection{Thermodynamic and magnetic quantities in the coupling files} \label{sec:interface}

\begin{figure}[h!]
    \centering
    \includegraphics[width=0.5\textwidth]{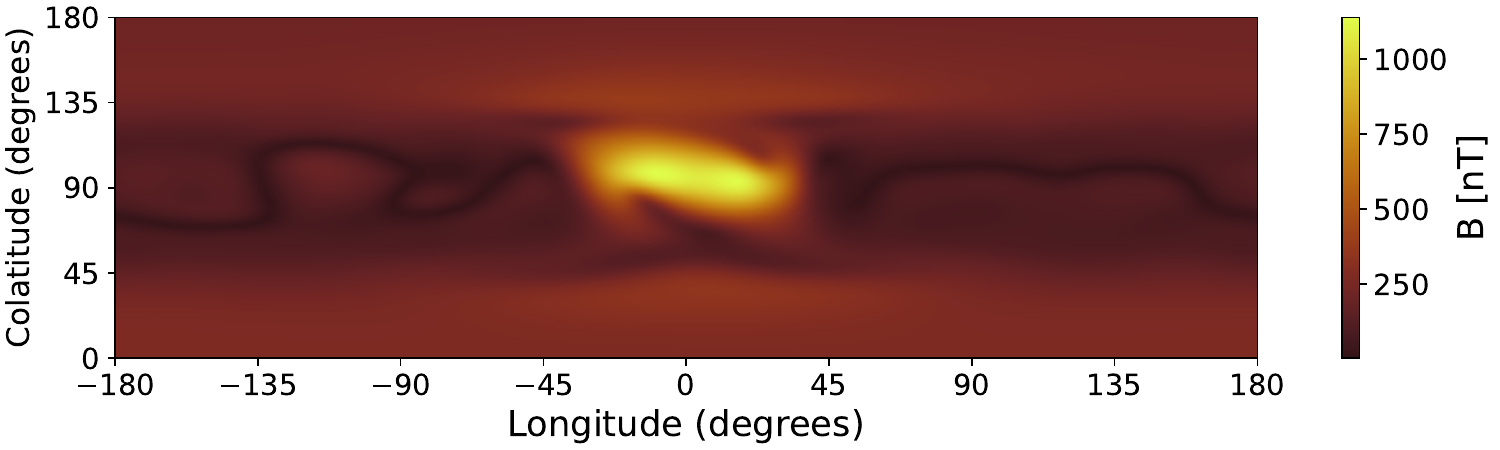}
    \caption{2D surface map of the magnetic field amplitude as saved in a coupling file. The magnetic field amplitude is computed using the three magnetic field components $B_{r}$, $B_{lon}$, $B_{clt}$ saved in the 138th ASCII file used for the coupling between COCONUT and EUHFORIA for the case named "TDm\_3". The 138th file, marking the time $t=5.5$ hours, captures when the flux rope starts crossing the surface at $R_{b}=21.5~\;R_{\odot}$.}
    \label{fig:Binterface}
\end{figure}

As detailed in Sections~\ref{sec:outputprep} and \ref{sec:updated euhforia}, the three components of the magnetic field, of the velocity as well as the temperature, and the density at the surface $R_{b}=21.5~\;R_{\odot}$ are regularly saved to files. This series of files is used as the inner boundary for EUHFORIA.

Figure~\ref{fig:Binterface} shows the magnetic field amplitude computed using the magnetic field components, which are saved in the coupling file at the moment when the flux rope, "TDm\_3", begins crossing the $R_{b}=21.5~\;R_{\odot}$ surface of COCONUT. This figure shows a weak magnetic field intensity line corresponding to the equatorial streamer belt. We also observe that the increase in the magnetic field associated with the arrival of the flux rope is mainly located at a colatitude of $90^\circ$ and a longitude of $0^\circ$. Since a 180-degree rotation was applied during the creation of the file (cf.\ Sect.~\ref{sec:outputprep}), we can deduce that the propagation direction remains aligned with the active region of implementation. No major deflection of the CME was observed during its propagation in COCONUT. This could result from our choice to use a solar minimum case for the background solar wind.

\begin{figure}[h!]
    \centering
    \includegraphics[width=0.45\textwidth]{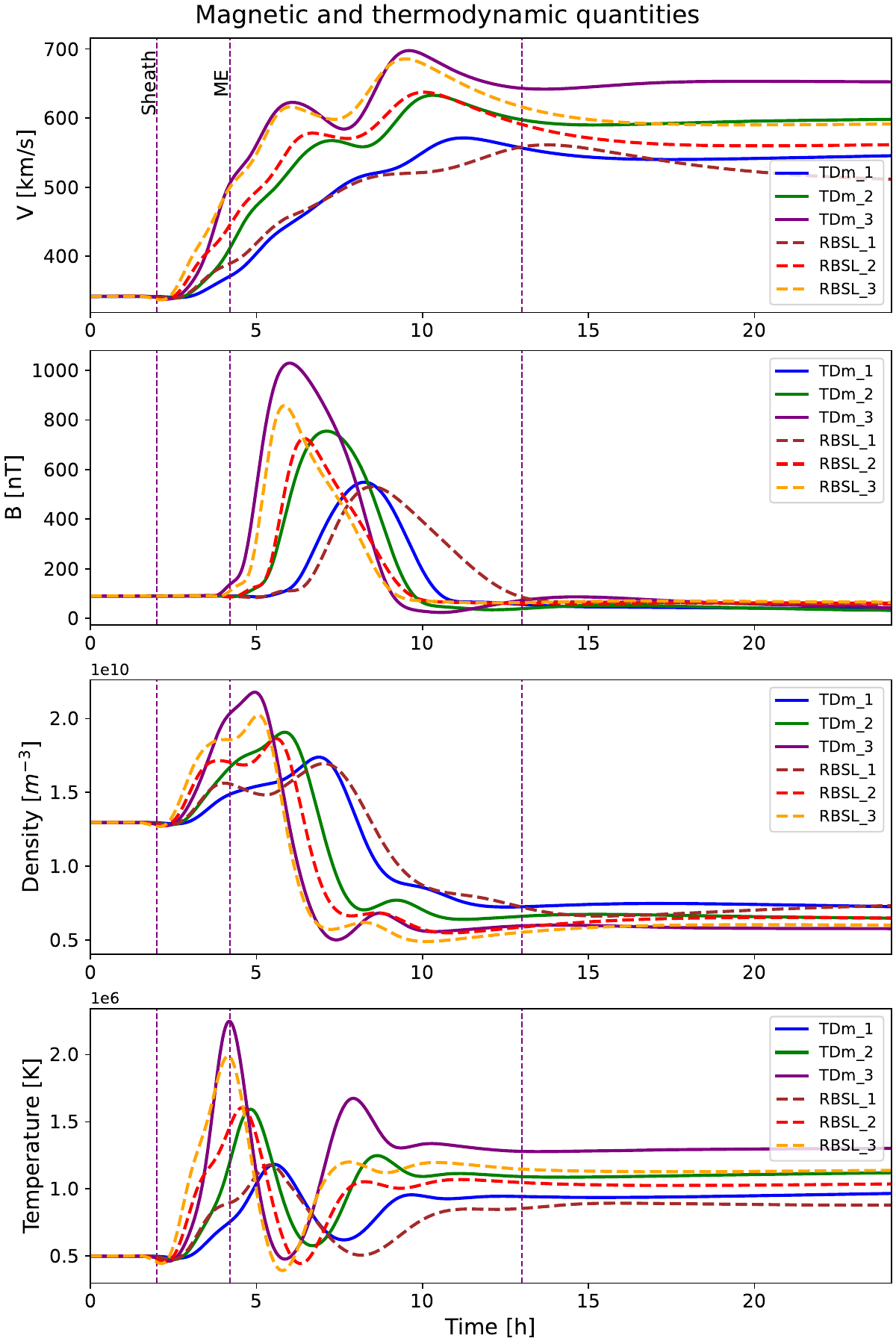}
    \caption{Velocity, magnetic field, density, and temperature evolution as a function of time, starting from the magnetogram date. Each line corresponds to a different simulation in the background solar wind. These profiles are consistent with the propagation of a flux rope with a sheath ahead. The purple vertical lines approximately mark the start of the sheath, the beginning of the magnetic ejecta (ME), and the end of the ME for the simulation named "TDM\_3.} 
    \label{fig:interface}
\end{figure}

To understand the data transferred to EUHFORIA, we read all 600 coupling files for all six simulations. We extract the values at a colatitude of $90^\circ$ and a longitude of $0^\circ$ and plot the temporal evolution of various global quantities in Fig.~\ref{fig:interface}: the magnetic field amplitude $B$, the density $n$, the velocity amplitude $v$, and the temperature $T$. It is important to note that in EUHFORIA, as described in Section~\ref{sec:boundary}, the inner boundary rotates at each time step by an angle equal to $t\Omega$. This means that the longitude of $0^\circ$ in the boundary map is perfectly aligned with Earth only at the initial time $t=0$ (the magnetogram date). Since the solar rotation rate is slow, Fig.~\ref{fig:interface} still provides a good estimation of the perturbation transferred towards Earth.

Figure~\ref{fig:interface} shows that the different simulations have quite similar profiles. The main differences between all the cases are the amplitude of the peaks reached and the duration of the disturbances generated. In detail, the arrival of the disturbance caused by the propagation of the flux ropes increases all quantities. The first to reach a local maximum is the temperature. At the peak of the temperature, we notice a slight change in the slope of the total magnetic field, $|B|$. According to \citet{Linan23}, this could indicate the transition between the sheath and the magnetic ejecta. During the crossing of the magnetic ejecta, the temperature decreases and then increases in the wake of the CME. After the flux ropes have passed, the temperature is higher than the initial moment due to constant heating, as explained in the previous section (cf.\ Sect.~\ref{sec:propagationincoco}).

The magnetic field profile shows a single increase before returning to its initial value once the magnetic ejecta has finished crossing the boundary $R_{b}=21.5~\;R_{\odot}$. For a given CME model (RBSL or TDm), the higher the initial magnetic flux, the higher the maximum value reached since the flux is directly proportional to the strength of the magnetic field within the flux rope.

Similarly, the velocity peaks in Fig.~\ref{fig:interface} are proportional to the initial magnetic flux. The higher the magnetic flux, the higher the velocity profile values reached. It should be noted, however, that for each simulation, the maximum reached is never more than $50\%$ of the initial speed listed in Table~\ref{tab:case}. This speed reduction can be attributed to the drag force exerted by the ambient solar wind, which acts to restrain the expanding flux rope. Such a force accounts for the gradual deceleration of CMEs as they travel through space \citep{Sachdeva15,Sachdeva17}. As described in Section~\ref{sec:propagationincoco}, the two local maxima observed in the velocity profile can be attributed to different phenomena. The first peak corresponds to the passage of the bulk of the CME, while the second is due to the acceleration created by the reconnection outflow in the wake of the CME \citep{Shibata01}.

Regarding the density in Figure~\ref{fig:interface}, the amplitude of the peaks in the density profiles is directly related to the speed of the CMEs: faster CMEs generate higher peak densities. This is because faster CMEs compress the plasma in front of them more effectively, leading to a denser, more compact region \citep{Winslow15}. Indeed, the fastest CMEs, "TDm\_3" and "RBSL\_3", show the highest peak. Conversely, the slower CMEs, "TDm\_1" and "RBSL\_1", exhibit lower peak densities due to less effective compression. After the passage of the CME, the density is lower than before. This can be attributed to the fact that the CME has swept up and removed a significant amount of solar wind plasma from its path, creating a rarefied region behind it. 

Finally, the patterns observed in Figure~\ref{fig:interface} are consistent with the observations made in the COCONUT domain described in Section~\ref{sec:propagationincoco}. Therefore, we can assume that the interpolation of the unstructured grid data of COCONUT, which was necessary to write the boundary file, had little impact on the transferred physics.

\subsection{Propagation continuation in EUHFORIA} \label{sec:propagationEUHFORIA}

\begin{figure*}[ht!]
    \centering
    \includegraphics[width=\textwidth]{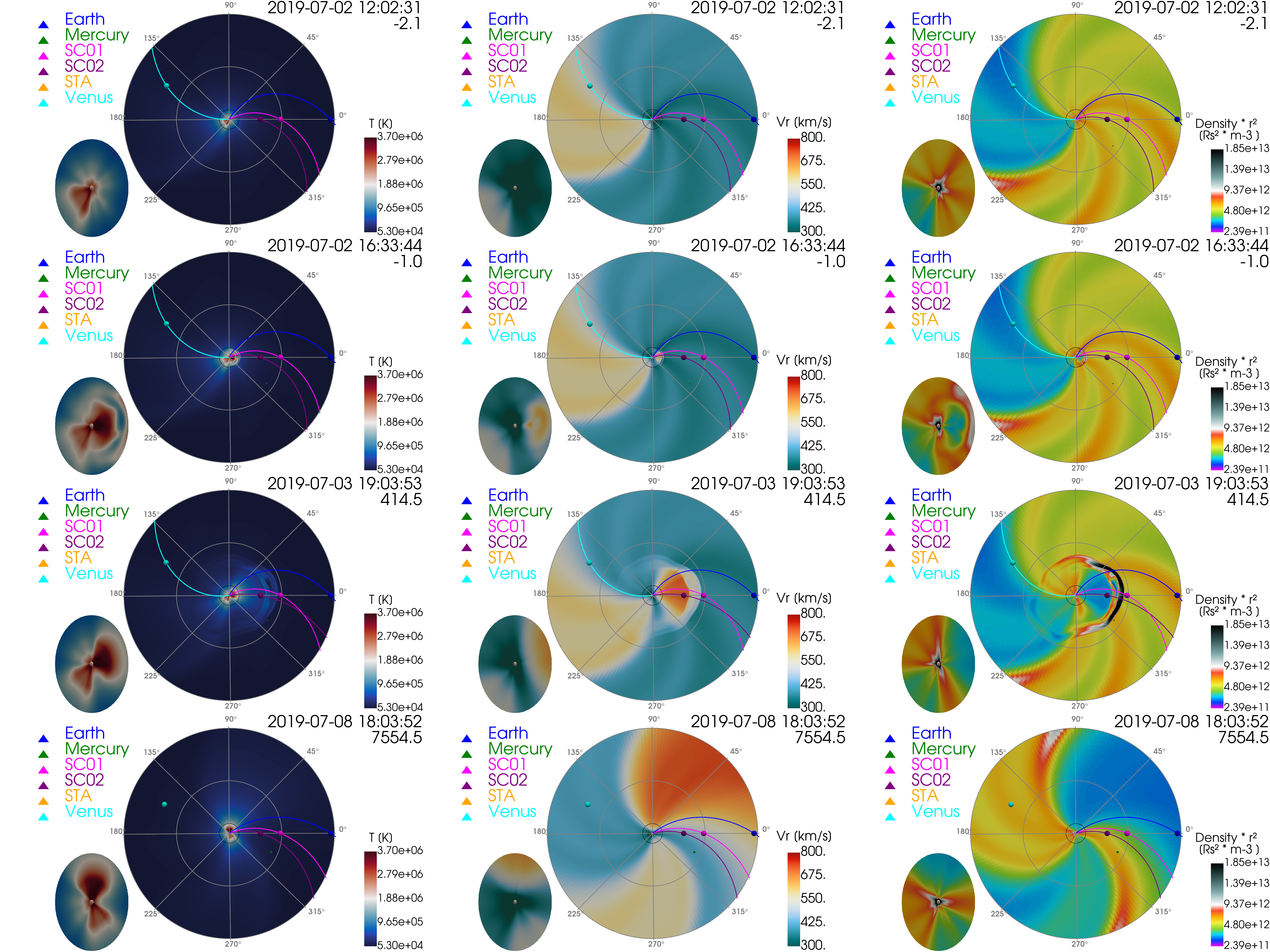}
    \caption{Composite Visualization of the Equatorial Plane from COCONUT and EUHFORIA. Each panel contains two parts: the inner disk is delineated by a black circle derived from COCONUT's output, while the rest is obtained from EUHFORIA. The region covered by COCONUT is also shown in the bottom left corner of each panel. The left panels represent the temperature distribution in Kelvin,  the middle panels represent the radial velocity distribution in km/s, and the right panels represent the density multiplied by the square of the radius. From top to bottom, various time snapshots from EUHFORIA are presented. The number below the date in the top right of each panel indicates the time difference in minutes between the EUHFORIA and COCONUT timestamps. Each panel includes markers at the positions of different planets, virtual satellites, and the Parker spirals connecting them to the Sun.}
    \label{fig:cocoria}
\end{figure*}

Once the COCONUT simulations are complete and the coupling files generated, these files are used in EUHFORIA as boundary conditions described in Section~\ref{sec:updated euhforia}. For our EUHFORIA run, eleven days are used for the relaxation phase and seven days for the forecast phase. As a result, there are enough coupling files to cover only 1/7 of the forecast phase. During the remaining six days, the last available file is used. 

Figure~\ref{fig:cocoria} shows the equatorial distribution of temperature, radial velocity, and density in COCONUT and EUHFORIA at various times. Due to differing time steps in COCONUT and EUHFORIA, there is no exact temporal correspondence between the outputs of COCONUT and those of EUHFORIA. Consequently, the difference in time between the Paraview file produced by EUHFORIA and those made by COCONUT is noted in minutes in the top right corner of each panel. The first black circle marks the boundary of the COCONUT domain.

The top three panels of Figure~\ref{fig:cocoria} display the equatorial plane after the relaxation phase in EUHFORIA for the case named "TDM\_3". Throughout the 11-day relaxation period, the same coupling file is repeatedly read in EUHFORIA to create the solar wind. In Figure~\ref{fig:cocoria}, it is apparent that for the three variables ($n$, $v_{r}$, $T$), the EUHFORIA domain directly extends from the COCONUT domain. The solar wind maintains a coherent structure across both simulations, and the transition between COCONUT and EUHFORIA is seamless, showing no significant changes in the amplitude of these three variables. The information in the coupling file appears to be fully integrated into EUHFORIA. Notably, a high-speed stream opposite Earth is clearly distinguishable in the simulation. As previously discussed (cf.\ Sect.~\ref{sec:propagationincoco}), this is due to intense heating in the solar corona.

We can conclude that the steady-state solar wind in EUHFORIA can be constructed from the COCONUT interface at $R_{b}=21.5~\;R_{\odot}$.

After the relaxation phase, and for about one day in simulation time, the inner boundary of EUHFORIA is updated through temporal interpolation between the two nearest coupling files to the time $t$ being processed. The second row in Figure~\ref{fig:cocoria} captures the approximate moment when the sheath begins to enter EUHFORIA, occurring four and a half hours after the relaxation ends. Since the disturbance caused by the flux rope propagation has just reached the $R_{b}=21.5~\;R_{\odot}$ surface, the solar wind in EUHFORIA has only slightly evolved during these four hours. It is particularly noteworthy that despite reading different boundary files and performing temporal interpolations up to this point, no major and apparent numerical artefacts significant enough to impact the large-scale structure of the solar wind have been generated within the domain. For the other quantities saved in the coupling files and not represented in Figure~\ref{fig:cocoria} (such as the magnetic field components), a smooth transition at the interface between COCONUT and EUHFORIA can also be observed.

\begin{figure}[h!]
    \centering
    \includegraphics[width=0.5\textwidth]{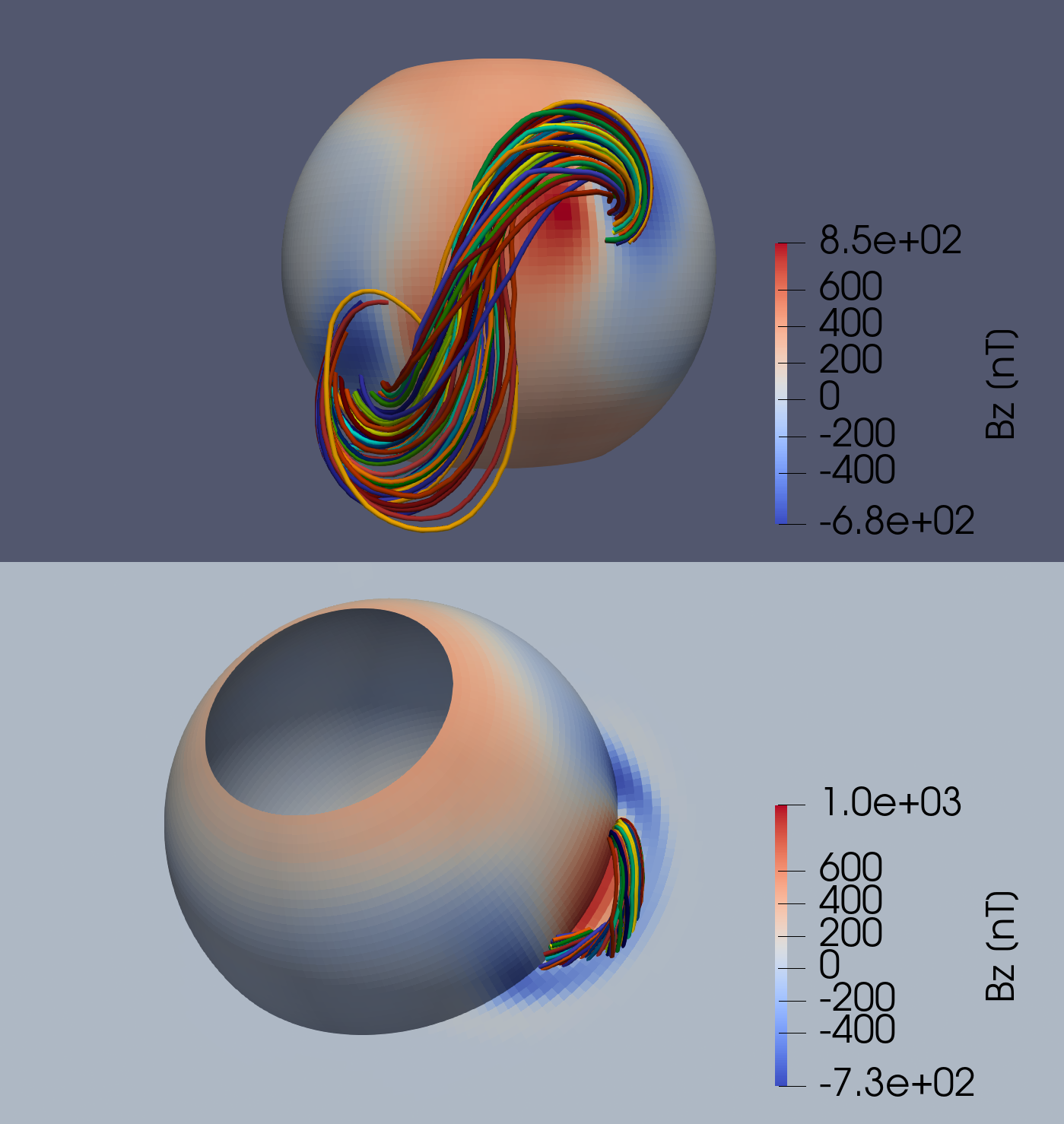}
    \caption{Visualization of the TDm flux rope model and the RBSL model in EUHFORIA. The upper panel shows the CME model for the case named the case named "RBSL\_3", while the bottom panel shows the simulation with the case named "TDm\_3". The coloured lines represent a sample of magnetic field lines from the flux ropes. The 3D sphere represents the inner boundary of EUHFORIA at $R_{b}=21.5R_{\odot}$. For the TDm simulation, the equatorial plane shows the $B_{z}$ magnetic field component in nT. }
    \label{fig:tdm_EUHFORIA}
\end{figure}

As a further demonstration that information is effectively transferred from COCONUT to EUHFORIA during the reading of boundary files, Figure~\ref{fig:tdm_EUHFORIA} shows the magnetic field lines of the "RBSL\_3" and "TDm\_3" flux ropes shortly after they cross the interface at $R_{b}=21.5~\;R_{\odot}$. The depicted field lines correspond to the nose of the CMEs, which is corroborated by the distribution of the $B_{z}$ component of the magnetic field that exhibits a sign change, as expected in the Titov-Démoulin model \citep{Linan23}. As in COCONUT (cf.\ Sect.~\ref{sec:propagationincoco}), the TDm flux rope is mainly in the equatorial plane, while the RBSL presents an "S-shape" as initially implemented. We conclude that the morphology of the CME is well transferred from one simulation to the other.

In EUHFORIA, the coupling files are read and used as boundary conditions as long as the simulation time is less than the date of the last available files, corresponding to about one day in simulation time. Afterwards, the previous available boundary file is utilized until the simulation concludes. The third row in Figure~\ref{fig:cocoria} shows the equatorial plane approximately 6 hours after the date corresponding to the last coupling file. By this time, the disturbance has propagated toward Earth with radial expansion. Between the virtual satellites labelled 'SC01' and 'SC02,' a high temperature and density region is observed. Based on COCONUT observations, we identify this region as the sheath, which continues to develop in EUHFORIA. As the flux rope moves forward, material that cannot escape accumulates upstream of the propagation. Like in COCONUT, a higher speed peak is not located in this region but in the wake of the disturbance.

Although the disturbance is primarily visible in the direction of propagation (the X-direction), a circular band of high density covering the entire domain is discernible. We suggest that this is due to a shock created upstream of the flux rope in all directions from the onset of the COCONUT simulation when the CME begins its evolution at speeds exceeding 1700 km/s.

Finally, at the very end of the simulation, as shown in the last row of Figure~\ref{fig:cocoria}, the solar wind has drastically changed from the initial conditions. Nevertheless, the high-speed stream present after the relaxation phase has shifted according to the co-rotated frame. Another high-speed stream with greater amplitude is also noticeable in the positive X-direction. As explained in Section~\ref{sec:propagationincoco}, the velocity in the wake of the CME never decreases due to the continuous heating associated with maintaining the active region that originated the flux rope. This rapid plasma flow is constantly transferred into EUHFORIA, resulting in a high-speed region. It is also important to note that the density in this area is lower than initially because the flux rope has pushed material ahead, creating a cavity in its wake (cf.\ Sect.~\ref{sec:propagationincoco}).
\subsection{Profiles at Earth} \label{sec:Earth}

\begin{figure}[h!]
    \centering
    \includegraphics[width=0.5\textwidth]{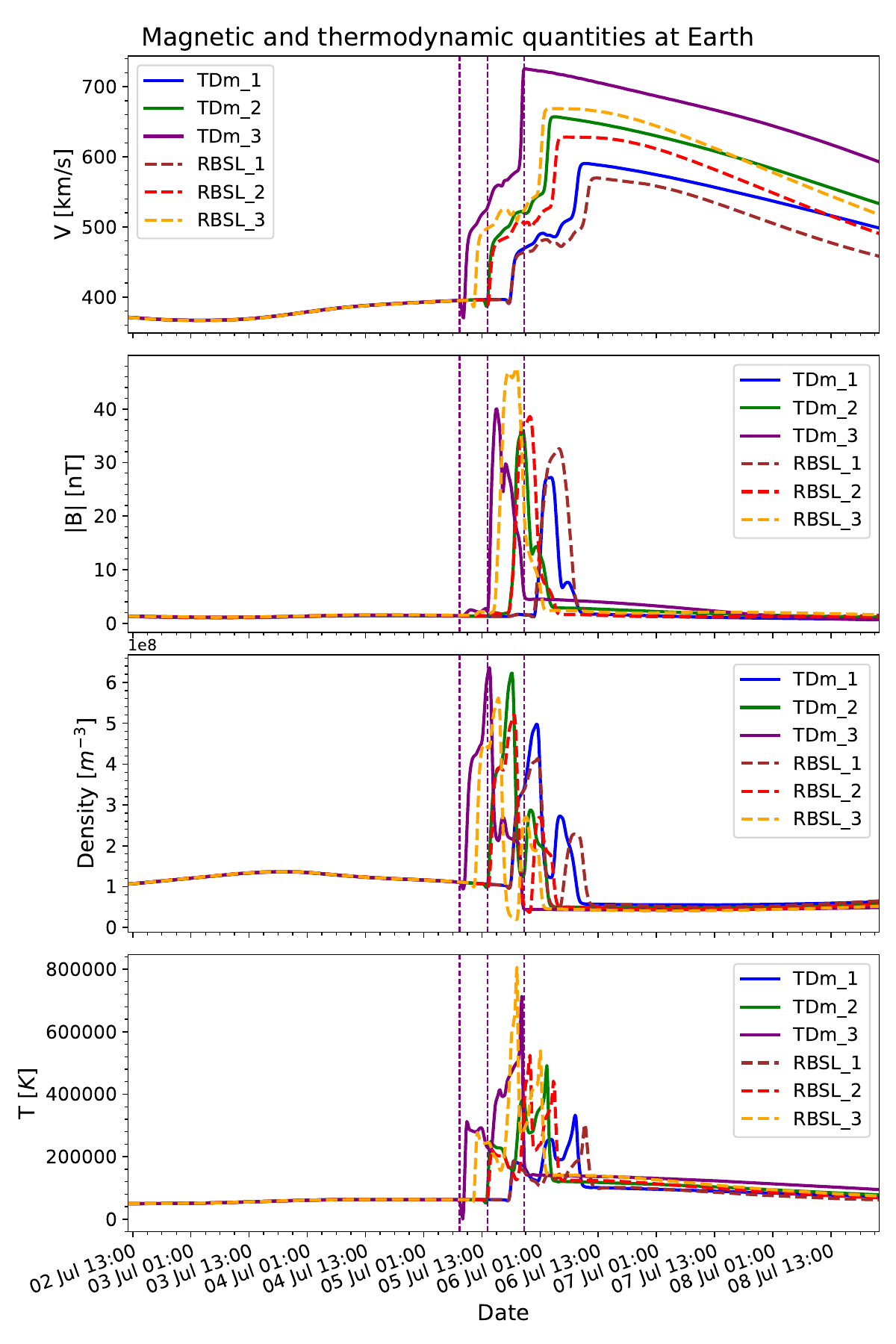}
    \caption{Evolution of Earth's magnetic and thermodynamic quantities in EUHFORIA. From top to bottom, the figure shows the profiles of the total velocity in km/s, the total magnetic field in nT, the density in $m^{-3}$, and the temperature at Earth as predicted by EUHFORIA. Each coloured line represents a different simulation, as in Fig.~\ref{fig:interface} and the purple vertical lines approximately mark the start of the sheath, the beginning of the ME, and the end of the ME for the simulation named "TDM\_3".}
    \label{fig:Earth}
\end{figure}

EUHFORIA is used mainly to predict the evolution of a CME's thermodynamic and magnetic properties at Earth. Consequently, Figure~\ref{fig:Earth} shows the evolution of the total velocity, total magnetic field, density, and temperature at Earth for the different simulations performed in this study.

First, in Figure~\ref{fig:Earth}, we observe that the amplitude of the different profiles depends on the simulation, as noted at the interface between the two simulations (Sect. \ref{sec:interface} and Fig.~\ref{fig:interface}). For example, the initially fastest flux ropes, i.e., the cases $TDm\_3$ and $RBSL\_3$, have the highest peak velocities at Earth. Additionally, we notice that the peak values are of the same order of magnitude as those observed at the interface (cf.\ Fig.~\ref{fig:interface}), indicating that the flux ropes undergo significant deceleration mainly in the solar corona, modelled here by COCONUT. Regarding the velocity profile, the maximum is preceded by a sharp increase in each simulation. In COCONUT, the velocity increase was smoother. However, this may be due to the spatial resolution used. For Figure~\ref{fig:Earth}, the values are obtained every 10 minutes, while for Figure~\ref{fig:interface}, the data points are approximately 2.4 minutes apart.

Similarly to the velocity, the total magnetic field profiles' amplitudes follow the same hierarchy described at the interface between COCONUT and EUHFORIA (cf.\ Sect.~\ref{sec:interface}). The higher the initial magnetic flux of the flux ropes, the higher the amplitude at Earth.

The density and temperature profiles exhibit complex patterns indicating the passage of multiple structures at Earth. Despite a noticeable increase due to the sheath passage, the maximum density ranges between $6 \times 10^{10} \text{m}^{-3}$ and $4 \times 10^{8} \text{m}^{-3}$, which is nearly two orders of magnitude smaller than the maximum values obtained in COCONUT at $R_{b}=21.5R_{\odot}$, which vary between $1.6 \times 10^{10} \text{m}^{-3}$ and $2.2 \times 10^{10} \text{m}^{-3}$ (cf.\ Fig.~\ref{sec:interface}). The dispersion of matter ahead of a CME as the magnetic structure extends is observed both in the solar corona and the heliosphere. For instance, using a sample of 45 ICMEs observed by Helios 1/2 and the Parker Solar Probe, \citet{Temmer22} found that the region in front of the magnetic driver becomes denser than the ambient solar wind at about 0.06 AU, with density decreasing as a linear function of distance. A drop in temperature also accompanies this reduction in density.

After the passage of magnetic and thermodynamic perturbations, around July 6 at 10 AM, the temperature is higher than before the event, and the density is lower, consistent with observations made in COCONUT. This consistency again demonstrates a coherence between the changes in the solar wind in the solar corona modelled by COCONUT and those occurring in the heliosphere modelled by EUHFORIA. However, more detailed analyses, which are beyond the scope of this study, are needed to fully explain patterns specific to the evolution in EUHFORIA, such as the sawtooth structure of the temperature and the significant increase in density after the passage of the magnetic structure.

\begin{figure*}[ht!]
    \centering
    \includegraphics[width=\textwidth]{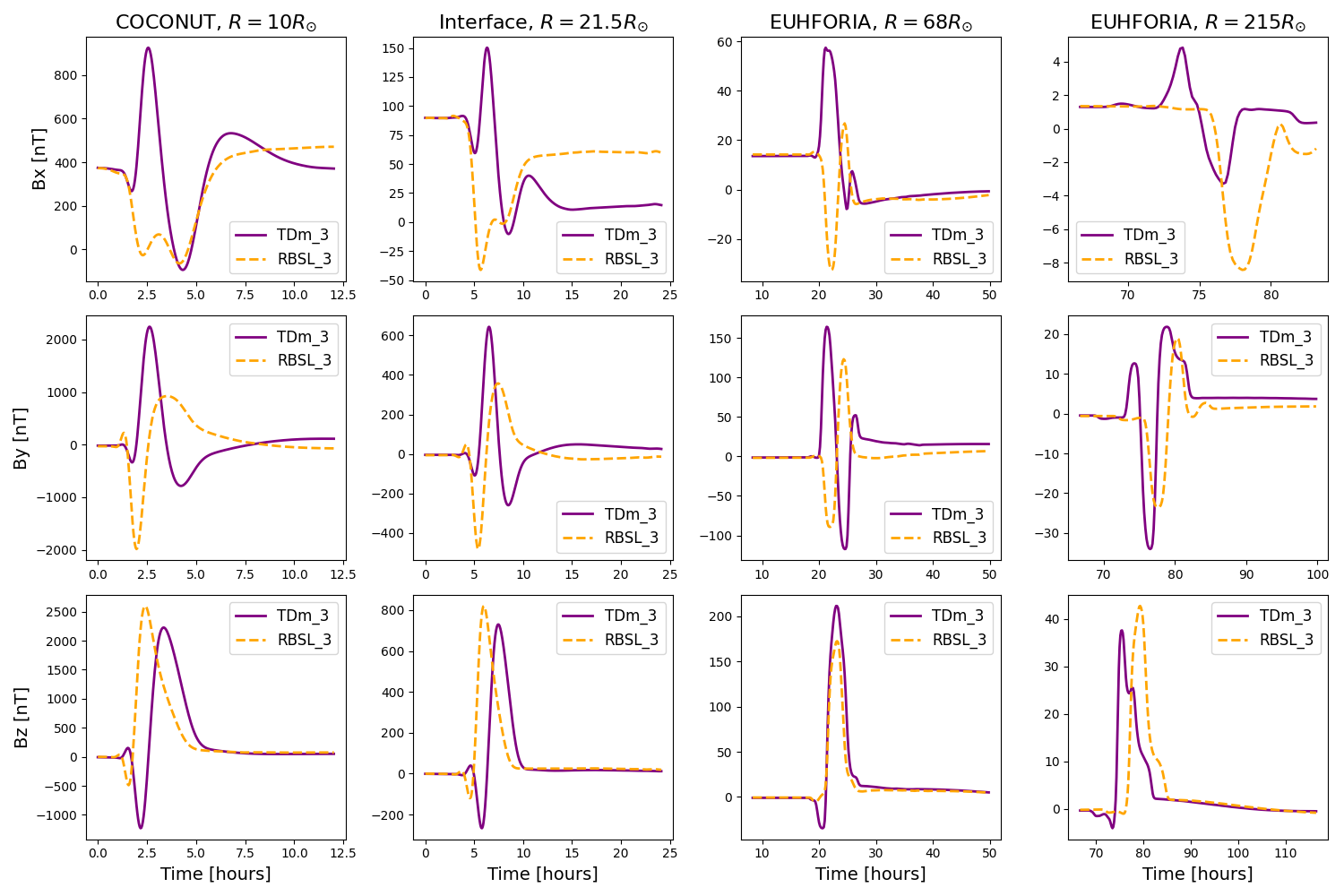}
    \caption{Evolution of the magnetic components at different distances from the Sun. The first row depicts the evolution of the $B_{x}$ component in nT, the second row shows the $B_{y}$ component, and the third row presents the $B_{z}$ component. The columns correspond to different distances from the Sun at a colatitude of $\theta=90^{\circ}$ and a longitude of $lon=0$. Specifically, the first column shows the evolution at $R=10~\;R_{\odot}$ in COCONUT, the second column at $R_{b}=21.5~\;R_{\odot}$, which is the interface between COCONUT and EUHFORIA, the third column at $R=68~\;R_{\odot}$ in EUHFORIA, and the final column illustrates the evolution at Earth.}
    \label{fig:earth_b}
\end{figure*}

So far, we have described only the evolution of the total magnetic field, $|B|$. However, the detailed distribution of the magnetic field is crucial since the strongest geomagnetic storms occur when the $B_z$ component of the magnetic field of the CME is opposite in direction to that of the magnetopause \citep{Lugaz16}. We placed virtual satellites in COCONUT and EUHFORIA at various distances along the propagation direction to track the evolution of the flux rope distributions. Specifically, we examine the time evolution of the magnetic field components, $B_{x}$, $B_{y}$, and $B_{z}$ at distances of $R=10~\;R_{\odot}$ in COCONUT, at the interface between the two simulations (by reading the coupling files), at $R=68~\;R_{\odot}$ in EUHFORIA, and at Earth. We aim to capture the arrival of the CME towards Earth by positioning the virtual observation points at a colatitude of $\theta=90^{\circ}$ and a longitude of $lon=0^{\circ}$. It's worth noting that COCONUT and EUHFORIA separately do not evolve in the same reference frame. To transition to the co-rotated frame of EUHFORIA, the 21.5 $R_{\odot}$ boundary of COCONUT is rotated at each time step (cf.~Sect.~\ref{sec:boundary}). This means that the virtual satellites in COCONUT and at the interface are slightly offset from those placed in EUHFORIA. Therefore, we do not expect to have perfectly identical profiles between COCONUT and EUHFORIA.

To enhance the figure's readability, we present only the flux ropes with the highest magnetic flux, the cases named "TDm\_3" and "RBSL\_3". The other instances exhibit similar profiles but with lower amplitudes.

The first column of Figure~\ref{fig:earth_b} shows the evolution of the magnetic field components at ten solar radii. This point is within the domain covered by COCONUT. First, we see that the $B_{x}$ component is not zero, indicating that the measurement point is slightly offset from the nose of the CME. The profiles of the three magnetic field components depend on the flux rope model used. For example, both models have a $B_{y}$ profile that exhibits symmetry with respect to the horizontal axis. The $B_{z}$ component follows a similar evolution in both simulations, with two sign changes. However, the TDm flux rope reaches negative values more than twice as small as the RBSL flux rope, whose negative bump is more discreet. In the simulation with the TDm flux rope model, the first positive bump in $B_{z}$, accompanied by a negative peak in the $B_{x}$ and $B_{y}$ components, corresponds to the magnetic field of the sheath according to \citet{Linan23}.

At the interface between the two simulations (see the second column in Fig.~\ref{fig:earth_b}), the profiles for the three magnetic field components are very similar to those obtained closer to the Sun. The main sign changes are present for the $B_{y}$ and $B_{z}$ components. According to \citet{Linan23, Guo24, Guo24b}, in a minimum activity solar wind in COCONUT, the flux ropes develop radially from the Sun to the outer boundary without deterioration of their overall structure. This explains why we expect to obtain similar profiles at different distances from the Sun in COCONUT.

In Fig.~\ref{fig:earth_b}, we can see that while the profile shapes are preserved, the amplitude of the curves has significantly decreased as we move away from the Sun's surface. For example, the $B_{z}$ maximum decreases from approximately 2226 nT at 10 $R_{\odot}$ to around 729 nT at 21.5 $R_{\odot}$, representing a reduction of about 67.25\%. The reduction in the magnetic field of a CME due to its radial expansion and interaction with the surrounding solar wind is a well-known phenomenon, thanks to in-situ observations and measurements made at different distances in the heliosphere by multiple spacecraft \citep[e.g.][]{Liu05, Winslow15}.

In EUHFORIA, at approximately 68 solar radii, the main patterns of the different magnetic field components are still clearly present, which is consistent with efficient coupling. Nevertheless, for the $RBSL_3$ simulation, the profile of the $B_{x}$ component shows some differences compared to the previous profile, such as a noticeable sign change occurring around 25 hours after the start of the simulation. A positive bump is also observed at the rear of the magnetic ejecta in the $B_{y}$ component, which was not seen at the interface and in COCONUT. Assuming that the differences are not due to poor coupling (which would have been evident in the previous section), profile differences could potentially indicate a slight deflection of the CME during its evolution in EUHFORIA \citep{Kay13, Kay15}. These differences are also a result of the frame discrepancy between COCONUT and EUHFORIA, which prevents us from having perfectly aligned virtual satellites. Moreover, it should be noted that we did not expect to find the same profiles as at the interface because, after travelling more than 40 solar radii, the flux rope has developed and interacted with the solar wind.

At Earth (cf.\ last column in Fig.~\ref{fig:earth_b}), the amplitude of the different profiles has drastically decreased compared to those obtained in COCONUT. We can perform a power law fit by taking the maximum magnetic field values at the four distances shown in Fig.~\ref{fig:earth_b}. We find that the decrease in the magnetic field follows a linear trend with a slope exponent of approximately $-1.50$ for both the RBSL and TDm cases. This slope is higher than the value reported by \citet{Winslow15}, who derived an exponent of $-1.89 \pm 0.14$ based on observational data, but it is closer to the one obtained by \citet{Liu05} ($-1.40 \pm 0.08$). However, it is important to note that our fit is highly approximate since we consider four data points. This limitation means the fit's precision and robustness are constrained, and any conclusions drawn should be interpreted with caution.

Once again, the profiles in EUHFORIA at Earth are consistent with those obtained closer to the Sun. The characteristic pattern in the evolution of the $B_{y}$ component, notably, is observed at all distances from the Sun. Indeed, the orange curve (RBSL\_3) shows a decrease followed by a pronounced positive peak before returning to its initial value. This significant variation is consistently evident across different distances, illustrating the coherent propagation of the magnetic field structure throughout the corona and the heliosphere.

Despite the common patterns observed at previous distances, the profile at Earth shows notable changes compared to those obtained at a distance of 68 solar radii ($R= 68R_{\odot}$). Being far from the inner boundary in EUHFORIA, these differences are not due to coupling issues but rather to the dynamics of traversing the heliosphere. The most significant change is the drastic decrease in the amplitude of the first positive peak of the $B_{y}$ component in the case named TDm\_3, dropping from around 157 nT to 12 nT, with the local minimum shifting from -110 nT to -35 nT. As mentioned earlier, this could be due to a deflection of the CME in the heliosphere, causing Earth to traverse the flank of the CME rather than its nose. Another contributing factor could be the erosion of the CME’s magnetic structure due to interactions with the ambient solar wind.

This erosion results from magnetic reconnection between the CME’s internal magnetic field and the interplanetary magnetic field (IMF) \citep{Hosteaux21}. However, as the CME travels through the heliosphere, it encounters a solar wind whose polarity and magnetic field components vary significantly. This variability means that magnetic reconnection is not uniform throughout the CME structure. Consequently, reconnection rates and the extent of erosion are influenced by the spatially and temporally fluctuating conditions along the CME’s path, causing certain areas of the CME to undergo more erosion than others.

As a result, the erosion process is complex, varying both spatially across the CME and temporally as it progresses through different solar wind conditions. The observed changes in the magnetic field components at different distances in the heliosphere could also potentially be due to a rotation of the CME structures \citep[e.g.][]{Regnault23,Ma24}. We attempted to highlight this rotation by tracing the magnetic field lines associated with the CME at various stages of the simulation, similar to the approach presented in Fig.~\ref{fig:tdm_EUHFORIA}. However, we could not clearly demonstrate a distinct rotation within the magnetic structure in either the TDm or RBSL simulations. Currently, limitations in spatial resolution make it difficult to conclusively identify rotation in these structures, and future simulations with higher resolution may improve our ability to capture this phenomenon more clearly.

Finally, both rotation and erosion are complex processes that evolve as the CME encounters different solar wind conditions. Further analysis is needed to fully understand the interactions between the solar wind and the CME, including the CME’s deflection, rotation, and spatially non-uniform erosion, all of which impact its evolution through the heliosphere in EUHFORIA. Such an in-depth analysis, however, lies beyond the scope of this paper.

\section{Summary and conclusion} \label{sec:conclusion}

In the heliospheric simulations with EUHFORIA, CME models are inserted at the inner boundary at $R_{b}=21.5~\;R_{\odot}$. This insertion is done entirely independently of the coronal model used to construct the solar wind. As a result, we miss all the CME dynamics and evolution in the solar corona (cf.\ Sect.~\ref{sec:Vanilla EUHFORIA}). To address this, we present the first time-dependent coupling between the coronal simulation with COCONUT and the 3D space weather forecasting simulation with EUHFORIA.

Using a specific magnetogram as input, we performed a set of six runs in COCONUT, where a flux rope is implemented at the solar surface. In three cases, the flux rope was modelled with the TDm CME model, while in the other three, we used the RBSL model (cf.\ Sect.~\ref{sec:testcases}). For this study, we used the new full MHD formalism developed by \citet{Baratashvili24} and detailed in Section~\ref{sec:fullmhd}. At regular intervals, the three components of the magnetic field, the three components of the velocity, the temperature, and the density on the 2D surface at $R_{b}=21.5~\;R_{\odot}$ are saved in a set of coupling files.

We modified the EUHFORIA code so that the values of the thermodynamic variables and magnetic components at the inner boundary $R_{b}=21.5~\;R_{\odot}$ are updated at each time step by reading the created boundary files (cf.\ Sect.~\ref{sec:updated euhforia}). After briefly presenting the propagation of the flux ropes in COCONUT and the impact of the full MHD formalism on the properties of the solar wind (cf.\ Sect.~\ref{sec:propagationincoco}), we focused on how the simulated coronal data are transmitted into EUHFORIA (cf.\ Sect.~\ref{sec:propagationEUHFORIA}). We particularly examined the time profiles at Earth generated by the arrival of CMEs initially implemented in COCONUT (cf.\ Sect.~\ref{sec:Earth}). Among the various results, our main findings are : 

\begin{itemize}
\item The full MHD formalism, used for the first time for time-dependent COCONUT runs in a minimum activity solar wind background, does not disrupt the propagation of the initially implemented flux ropes. As in the polytropic version of COCONUT \citep{Linan23,Guo24}, we observe a self-consistent radial expansion of the flux rope with matter accumulating upstream and the pinching of the CME legs below the core. However, future studies are needed to validate the time-dependent full MHD formalism in a maximum activity scenario.
\item Since the inner boundary of COCONUT remains unchanged throughout the simulation, the footpoints of the CME cause continuous heating of the solar corona. This results in a high-speed stream in the wake of the CME. In future studies, we suggest modifying COCONUT to allow for the temporal evolution of the solar surface, for instance, using a series of magnetograms. The goal is to gradually dissipate the active region created by the initial addition of the flux rope.
\item In EUHFORIA, the same coupling file from COCONUT is used to initially create the solar wind after a relaxation process. We find that the solar wind thus developed in EUHFORIA is a direct extension of the solar wind created in COCONUT. The transition between the two simulations is smooth, indicating that the creation and reading of the boundary file do not alter the information transmitted by COCONUT.
\item The initial perturbations caused by the propagation of a flux rope in COCONUT, visible as significant increases in speed, temperature, and density, continue to develop in EUHFORIA. This validates the time-dependent coupling between the two simulations. Moreover, in the standard version of EUHFORIA, a possible sheath would only develop from 0.1 AU onwards since the coronal model used was not affected by the propagation of a CME. In various studies using EUHFORIA, the size and amplitude of the sheath are often underestimated compared to observational expectations. However, with the coupling to COCONUT, a pre-formed sheath is inserted at the inner boundary. We suggest that this should better represent the sheath in the heliosphere, i.e., closer to observations.
\item The thermodynamic and magnetic profiles observed at Earth are consistent with those obtained at the interface between the two simulations, reinforcing the idea that simulation data are well transmitted from COCONUT to EUHFORIA. While the peak speed observed at Earth is close to that obtained at the interface of the two simulations, we find a significant decrease in the amplitude of temperature, density, and magnetic field. This result is consistent with various observations using multi-spacecraft in-situ measurements, indicating that the density of the region ahead of the flux rope decreases linearly with distance in the heliosphere \citep[e.g.][]{Liu05,Winslow15}. 
\item By examining the evolution of the three magnetic field components $B_{x}$, $B_{y}$, $B_{z}$ at different distances from the Sun, we found that, despite new dynamics due to interaction with the solar wind in the heliosphere, the main structural characteristics of the magnetic field are preserved as the CME propagates outward. This means that the properties of the flux rope, such as its morphology and magnetic flux, need to be carefully controlled at the Sun's surface in COCONUT since they directly impact the profiles observed at Earth and, consequently, on the geoeffectiveness of an event. The initial conditions set in the coronal model play a crucial role in determining the behaviour of the CME throughout its journey to Earth.
\end{itemize}

As a primary outcome of the reported implementation and validation efforts, we can conclude that the resulting new simulation chain COCONUT + EUHFORIA is a valuable tool for understanding the dynamics of CMEs from the Sun to the Earth. Now that the coupling has been introduced, the next logical step is to test the effectiveness of this simulation chain in predicting profiles generated by an actual event. For this kind of study, the initial properties of the flux ropes, such as their geometry and magnetic flux, should be derived from observations rather than arbitrarily set, as we have done in our work.

Several improvements can be made to ensure that the COCONUT + EUHFORIA toolchain is fully operational for space weather purposes. In particular, both simulations need to be able to run concurrently so that the outer boundary of COCONUT can be directly processed in EUHFORIA without waiting for the run to finish, similar to the workflows of the interactive tools CORHEL-CME \citep{Linker24}. Additionally, the parameters should be studied to identify settings (resolution, time step, etc.) that optimize computational time while maintaining a sufficiently good accuracy. In this study, we deliberately chose an excellent temporal resolution in COCONUT ($dt=0.005$ in code units) and a high spatial resolution in EUHFORIA. With this setup and the solver configuration used for the simulations, each COCONUT run lasted approximately one day and 19 hours. In contrast, an EUHFORIA run required 5 hours of wall time, plus several additional hours to create the coupling files from the text files generated by COCONUT.

Finally, improving the visualization tools available for displaying results would also be beneficial, and for this, we can rely on the figures presented in this study as a foundation.

Meanwhile, COCONUT will continue to be developed to enable an increasingly accurate representation of the solar corona that is closer to observations. We are implementing new heating profiles among the current development axes based on the density gradient perpendicular to the background magnetic field. We also intend to consider the heating associated with the wave turbulence-driven heating mechanism.

\begin{acknowledgements}
These results were obtained in the framework of the projects
C16/24/010 C1 project Internal Funds KU Leuven), G0B5823N and G002523N (WEAVE) (FWO-Vlaanderen),
4000145223 (SIDC Data Exploitation (SIDEX2), ESA Prodex), and Belspo project B2/191/P1/SWiM and the AFOSR basic research initiative project FA9550-18-1-0093.
For the computations, we used the VSC—Flemish Supercomputer Center infrastructure, which is funded by the Hercules Foundation and the Flemish Government department EWI.
LL acknowledges support from the ERC-2022-Advanced Grant 101095310 (TerraVirtualE, PI: Giovanni Lapenta). SP is funded by the European Union. Views and opinions expressed are, however, those of the author(s) only and do not necessarily reflect those of the European Union or [name of the granting authority]. Neither the European Union nor the granting authority can be held responsible for them. This project (Open SESAME) has received funding under the Horizon Europe programme (ERC-AdG agreement No 101141362).
\end{acknowledgements}

%
%


\bibliographystyle{aa} 
\bibliography{biblio.bib}

\begin{appendix}

\end{appendix}


\end{document}